\documentclass[aps,twocolumn,showpacs,superscriptaddress,preprintnumbers,nofootinbib,
amsmath,amssymb]{revtex4-1}

\usepackage{graphicx} 
\usepackage{tikz}
\usetikzlibrary{arrows}
\usepackage{multirow}
\usepackage[colorlinks=true,urlcolor=blue,linkcolor=blue,citecolor=blue]{hyperref}
\usepackage[normalem]{ulem}
\usepackage[capitalize]{cleveref}
\usepackage{amsfonts,amssymb}
 
\usepackage{comment}
\usepackage{flushend}

\usepackage[T1]{fontenc} 
\graphicspath{{fig/}}
\usepackage{mathtools}
\usepackage{xcolor}
\usepackage{graphicx}
\usepackage{subfigure}
\usepackage{dcolumn}
\usepackage{bm}

\definecolor{myblue}{RGB}{0, 100, 200}
\definecolor{myred}{RGB}{214, 39, 40}
\definecolor{mygreen}{RGB}{44, 160, 44}
\definecolor{mybrown}{RGB}{123, 64, 26}
\definecolor{mydarkblue}{RGB}{44, 77, 118} 

\begin{document} 

\preprint{KEK-QUP-2026-0012, KEK-TH-2856}

\title{Moving a Detector  to Probe New Neutrino Interactions: IWCD at Hyper-Kamiokande} 

\author{Thomas Schwemberger}
\email{tschwem2@post.kek.jp}
\affiliation{International Center for Quantum-field Measurement Systems for Studies of the Universe and Particles (QUP), High Energy Accelerator Research Organization (KEK), 1-1 Oho, Tsukuba, Ibaraki 305-0801, Japan}

\author{Volodymyr Takhistov}
\email{vtakhist@post.kek.jp}
\affiliation{International Center for Quantum-field Measurement Systems for Studies of the Universe and Particles (QUP), High Energy Accelerator Research Organization (KEK), 1-1 Oho, Tsukuba, Ibaraki 305-0801, Japan}
\affiliation{Theory Center, Institute of Particle and Nuclear Studies (IPNS), High Energy Accelerator Research Organization (KEK), 1-1 Oho, Tsukuba, Ibaraki 305-0801, Japan}
\affiliation{Graduate University for Advanced Studies (SOKENDAI), 1-1 Oho, Tsukuba, Ibaraki 305-0801, Japan}
\affiliation{Kavli Institute for the Physics and Mathematics of the Universe (WPI), UTIAS, The University of Tokyo, Kashiwa, Chiba 277-8583, Japan}

\begin{abstract}
Moving a detector through a beam with a spatially varying energy spectrum exposes the same target and apparatus to distinct incident spectra, enabling interaction spectroscopy. While movable neutrino detectors were put forth primarily to control systematic uncertainties, we show that detector motion enables probing the structure of fundamental interactions. We demonstrate this with the Intermediate Water Cherenkov Detector (IWCD), the movable detector of Hyper-Kamiokande in the J-PARC off-axis beam, considering neutrino non-standard interactions (NSI) as a benchmark. Exploiting ratios of neutral current to charged current event rates reduces common normalization uncertainties, while measurements at multiple off-axis positions can break degeneracies that persist for a single incident spectrum. Combining three off-axis positions and adopting a 5\% correlated normalization uncertainty benchmark, we project  95\% CL sensitivities to axial neutral current NSI of $-0.07 \lesssim \varepsilon^{uA}_{\mu\mu} \lesssim 0.06$ and vector NSI of $-0.10 \lesssim \varepsilon^{uV}_{\mu\mu} \lesssim 0.12$, as well as for charged current NSI of $-0.05 \lesssim \varepsilon^{udL}_{\mu\mu} \lesssim 0.05$. Axial NSI, which do not contribute to the ordinary matter potential, are complementary to neutrino oscillation and high energy scattering measurements. More broadly, detector motion provides a new way to distinguish interactions with different energy dependence.
\end{abstract}

\maketitle

\section{Introduction}

Changing a detector's position relative to an incident particle beam
can alter the energy spectrum it observes. For an off-axis neutrino beam, a
movable detector can therefore probe the same class of interactions under
different incident spectra. We show that this controlled spectral
variation provides a new handle on the structure of fundamental
interactions, allowing contributions from new physics that are degenerate
for a single spectrum to be distinguishable.

A broad range of theories beyond the Standard Model (SM) predict novel
neutrino interactions. New mediator fields, including additional gauge
bosons \cite{Farzan:2017xzy,Babu:2017olk,Heeck:2018nzc} and leptoquarks
\cite{Dorsner:2016wpm,Babu:2019mfe}, can induce additional couplings of neutrinos to quarks. For
momentum transfers well below the mediator mass, this can be effectively described by
four-fermion neutrino non-standard
interaction (NSI) operators  \cite{Wolfenstein:1977ue,Roulet:1991sm,Grossman:1995wx}.
Constraints on NSI originate from a variety of sources including studies of neutrino oscillations, coherent elastic
neutrino-nucleus scattering (CEvNS) and electroweak precision tests (see
Refs.~\cite{Ohlsson:2012kf,Farzan:2017xzy,Proceedings:2019qno} for
review). 

Controlled accelerator neutrino beams provide excellent environments to investigate these effects.
Such beams have long served as probes of the electroweak sector, from the
discovery of neutral current (NC) interactions~\cite{GargamelleNeutrino:1973jyy} to precision measurements of the weak
mixing angle~\cite{NuTeV:2001whx}. Long-baseline programs 
including T2K \cite{T2K:2011qtm} and NO$\nu$A \cite{NOvA:2007rmc}, as
well as the forthcoming Hyper-Kamiokande \cite{Hyper-Kamiokande:2018ofw}
and DUNE \cite{DUNE:2020lwj} experiments  enable precision
measurements of oscillation parameters with   intense beams and
high statistics detectors. Interpretations of long-baseline data have considered NSI
scenarios in which differences among preferred oscillation parameters are
reduced \cite{Chatterjee:2020kkm,Denton:2020uda,Chatterjee:2024kbn},
motivating complementary measurements that probe similar NSI effects
directly at  neutrino interaction vertices rather than through their
effects on propagation and oscillations.

Existing probes of new neutrino interactions are complementary but incomplete. Their sensitivity
depends on factors such as operator structure, target composition  and interaction
channel, and hence searches often leave blind regions in parameter space. Combining
measurements from different targets can lift the resulting degeneracies
\cite{Coloma:2017egw,Coloma:2022avw,Schwemberger:2023hee,Beatty:2025bgq},
but requires relating independently normalized datasets with different
experimental uncertainties. The oscillation matter potential arises from coherent forward scattering and in ordinary nonrelativistic unpolarized matter depends only on vector interactions. Consequently, some axial NSI combinations remain constrained only at around $\mathcal{O}(10^{-1})$~level~\cite{Abbaslu:2024jzo}. Direct scattering measurements probing NSI therefore provide an essential complementary probe. For event rate observables  measurements with a fixed incident spectrum can leave some NSI and normalization variations degenerate.

In this work we show that a movable neutrino detector enables the structure of new neutrino interactions to be resolved by measuring the same scattering processes with multiple controlled incident spectra.
Since the energy spectrum of an off-axis neutrino beam varies with beam angle, different detector positions expose the same target and apparatus to distinct spectra while preserving strong correlations among interaction rate normalizations and detector uncertainties across positions. Vector and axial contributions enter the interaction cross sections with different energy  and momentum transfer dependencies. Variations in their relative contributions thus  produce a characteristic position dependence in the observable rates. 

We illustrate our approach for the Intermediate Water Cherenkov Detector (IWCD), the movable near detector of Hyper-Kamiokande \cite{Hyper-Kamiokande:2018ofw}. IWCD implements the NuPRISM proposal to sample the Japan Proton Accelerator Research
Complex (J-PARC) neutrino beam over a range of off-axis angles  originally developed to study SM neutrino interactions and with capabilities to probe sterile neutrino oscillation effects~\cite{nuPRISM:2014mzw}. Related off-axis approaches have been
studied at DUNE-PRISM for beyond SM effects of sub-GeV dark sector  \cite{DeRomeri:2019kic} and for sterile neutrino and non-unitarity effects
\cite{Hernandez-Garcia:2026asl}. In contrast, here we consider detector motion as a spectroscopic probe of the structure of new neutrino interactions themselves. We demonstrate this using the ratio of NC and charged current (CC) event rates at multiple IWCD off-axis positions and show how these measurements can break degeneracies that remain for detectors with only a fixed incident spectrum.

This work is organized as follows. Sec.~\ref{sec:beam} introduces the J-PARC off-axis neutrino beam, the movable IWCD  and the event rate observables. In Sec.~\ref{sec:cross_sections} we overview the relevant interaction cross-sections and kinematic considerations. Sec.~\ref{sec:NSI}  introduces the NC and CC NSI framework. Sec.~\ref{sec:results} presents the statistical analysis and projected sensitivities. We conclude in Sec.~\ref{sec:conclusion}. Natural units are considered throughout.

\section{Off-Axis Beam and Movable IWCD}
\label{sec:beam}
 
Our strategy to exploit the movable detector to probe new physics effects relies on observations of the same target and
final states considering several distinct incident spectra. This is realizable with the J-PARC off-axis neutrino beam flux and the   mobility of the IWCD.

\subsection{Beam and Detector Configuration}

The Tokai to Hyper-K (T2HK) neutrino beam will be produced at J-PARC by directing 30 GeV protons onto a
graphite target.  The resulting charged hadrons, predominantly pions, are selected and focused by   magnetic horns before
decaying in flight in the decay volume~\cite{T2K:2012bge,Hyper-Kamiokande:2018ofw}.
The horn polarity produces either a neutrino-dominated forward
horn current (FHC) beam or an antineutrino-dominated reverse horn current (RHC)
beam.  We consider FHC operation throughout with the dominant beam
component of $\nu_\mu$.

The beam is to be delivered in spills containing eight bunches each with a full width of approximately 50 ns, distributed over a time interval of a few microseconds~\cite{Hyper-Kamiokande:2018ofw}. With an  $\mathcal{O}$(1~Hz) spill rate, this corresponds to an on-beam duty factor of $\ll 10^{-3}$ resulting in significant suppression of background processes unrelated to beam, including those associated with atmospheric neutrino interactions, using beam timing.

The IWCD is a planned water Cherenkov detector  
approximately 1~km downstream of the J-PARC neutrino source~\cite{nuPRISM:2014mzw,Hyper-Kamiokande:2018ofw}.  
Equipped with high resolution multi-PMT modules the detector will provide a fiducial mass of approximately 300 ton. 
The design enables the IWCD to move vertically within 50 m tall shaft, allowing it to sample off-axis angles from 1$^{\circ}$ to $4^{\circ}$  and observe neutrino spectra with peaks between approximately 0.4 GeV to 1 GeV.  Since the same detector is used at all positions the detector and
interaction uncertainties are correlated, while the flux changes with
off-axis angle. 

As benchmarks, we employ the expected neutrino fluxes computed in Ref.~\cite{nuPRISM:2014mzw,Naseby:2021olw} for off-axis angles at $1^\circ$, $2.5^\circ$ and $4^\circ$. At larger angles, the spectrum becomes narrower and shifts to lower
energies, peaking around
$1.0$, $0.6$ and $0.4$~GeV, respectively. Although the beam composition also changes with detector position  for our benchmark analysis we retain only the dominant $\nu_{\mu}$ component. For comparisons involving one, two  and three detector positions  we keep the total exposure fixed at a specific value of $N_{\rm POT}^{\rm tot}= 4.5\times 10^{20}$~\cite{nuPRISM:2014mzw} protons on target and divide it equally among considered detector positions (i.e. in case of three off-axis positions each  receives $N_{\rm POT}^{\rm tot}/3$ exposure).

\subsection{Data Samples and Observables}
\label{sec:samples}

At J-PARC neutrino beam energies below a few GeV, relevant for the IWCD, both NC and CC samples receive significant contributions
from quasi-elastic (QE) interactions, as well as multi-nucleon and resonant processes, with the latter
dominated by pion production~\cite{nuPRISM:2014mzw}.  
Events with visible pions can often be
distinguished from QE interactions, although final-state interactions
and pion absorption can cause resonant events to enter QE-like samples.
We focus on QE signal contributions.
Multinucleon, resonant  and other non-QE contributions are represented through   effective systematic effects and require dedicated detector-level analyses that we leave for future work. 

Significant event rates arise  from
charged-current quasielastic (CCQE) processes. Here we focus on $  
  \nu_\mu+n\to\mu^-+p $ with muons $\mu$, neutrons $n$ and protons $p$. 
This produces a muon-like Cherenkov ring in the experiment when the outgoing muon is above
the Cherenkov threshold, corresponding to a momentum
$p_\mu\simeq120$~MeV and energy $E_\mu\simeq160$~MeV in water.  For the dominant $\nu_\mu$ component, considering the FHC beam,
CCQE scattering occurs on the eight bound neutrons in
${}^{16}{\rm O}$.  In the case of $\bar{\nu}_{\mu}$, interactions $\bar{\nu}_{\mu} + p \to \mu^+ + n$ include hydrogen and thus ten target protons per water molecule.

For the analysis  we use the simulated flux spectra of Ref.~\cite{Naseby:2021olw}. To normalize these spectra that are provided in arbitrary normalization  at each benchmark off-axis angle we determine an angle-dependent flux normalization by matching the calculated CCQE interaction yield, before imposing the muon Cherenkov requirement, to the expected true CCQE event count reported in Tab.~III of Ref.~\cite{nuPRISM:2014mzw}. 
We associate the $1^\circ-2^\circ$, $2^\circ-3^\circ$ and $3^\circ-4^\circ$ angular intervals with the representative $1^\circ$, $2.5^\circ$ and $4^\circ$ flux spectra, respectively.
The same angle-dependent normalization  is then applied to NCQE. The Cherenkov threshold requirement is then applied to the CCQE sample for detection. Besides this restriction  the CCQE signal efficiency is taken to be unity.

For comparison, we consider NC quasi-elastic (NCQE) processes
\begin{align}
 \nu + {}^{16}{\rm O} &\to \nu+n+{}^{15}{\rm O}^{*}, \\
 \nu + {}^{16}{\rm O} &\to \nu+p+{}^{15}{\rm N}^{*}.\notag\
\end{align}
There is no outgoing charged lepton, while the ejected neutron is
neutral and the proton is typically below the Cherenkov threshold.
However, these interactions can nevertheless be detected through prompt
$\gamma$ rays from de-excitation of the residual nucleus and additional
$\gamma$ rays produced by final-state and secondary nucleon
interactions of the nucleus and surrounding water. Such NCQE-like
samples have been measured by Super-Kamiokande using atmospheric
neutrinos~\cite{Super-Kamiokande:2023oxd} and by T2K using accelerator
beam neutrinos~\cite{T2K:2019zqh}.

As a benchmark for the NCQE signal detection efficiency at IWCD we adopt  energy independent $80\%$, motivated by the T2K accelerator beam analysis\footnote{We note that this should not be interpreted as a strict IWCD detector performance indication, which requires a dedicated analysis beyond our scope.}~\cite{T2K:2019zqh}. For the adopted nuclear model around $16\%$ of events populate
the $(p_{1/2})^{-1}$ ground state and produce no primary
de-excitation $\gamma$ ray. Excited residual states can emit prompt
$\gamma$ rays, including the characteristic $6.18$~MeV from ${}^{15}{\rm O}^*$  and $6.32$~MeV from ${}^{15}{\rm N}^*$ lines, while
interactions of the ejected nucleon in the nucleus and water can produce
additional secondary $\gamma$ rays
~\cite{Ankowski:2011ei,T2K:2019zqh,T2K:2025ipr}. The absence of a
primary $\gamma$ ray thus does not imply that the event is necessarily undetectable. Exact signal efficiencies will be obtained from dedicated IWCD analyses.

Beam timing strongly suppresses backgrounds unrelated to the beam.  After timing cuts,
the leading backgrounds are expected to be beam correlated interaction channels and
their modeling, rather than radioactive or atmospheric events.  We do
not consider other NC channels and misidentified  components explicitly.

For a specified interaction sample at
off-axis position angle $\theta$, the expected event yield is
\begin{equation}
 N^\theta =
 T^\theta N
 \sum_{\alpha}
 \int dE_\nu 
 \phi^\theta_\alpha(E_\nu) 
 \sigma_{\alpha }(E_\nu) 
 \eta^\theta_{\alpha }(E_\nu) ,
 \label{eq:eventrate}
\end{equation}
where $T$ is the exposure, $E_{\nu}$ is neutrino energy, $N$ is the number
of interaction targets, $\sigma_{\alpha}$ is the interaction cross-section, $\eta^\theta_{\alpha}$ denotes selection efficiency and $\alpha$ labels the beam
 neutrino or antineutrino component.  

To analyze new physics effects, we are interested in distinct contributions to processes at different energies.
A convenient quantity to consider is the total flux-weighted ratio
\begin{equation}
 R^\theta =
 \frac{N^\theta_{\rm NCQE}}
      {N^\theta_{\rm CCQE}}
 \label{eq:angular_ratio}
\end{equation}
at each off-axis position. Considering $\nu_\mu$ as the dominant contribution, the common
exposure and leading flux normalization largely
cancel in this ratio. However, the energy dependence remains
\begin{equation} \label{eq:rate}
 R^\theta \sim
 \frac{\int dE_\nu 
       \phi^\theta_{\nu_\mu}(E_\nu)
       \sigma_{\rm NCQE}(E_\nu)
       \eta^\theta_{\rm NCQE}(E_\nu)}
      {\int dE_\nu 
       \phi^\theta_{\nu_\mu}(E_\nu)
       \sigma_{\rm CCQE}(E_\nu)
       \eta^\theta_{\rm CCQE}(E_\nu)} .
\end{equation}
Consequently, moving the IWCD changes the energy weighting of the NC and
CC cross-sections producing a position-dependent observable.
New interactions that modify their relative strength or energy
dependence can therefore change the pattern of $R^\theta$ across
positions, even when their effect at one position is degenerate with
an overall normalization.  Since the outgoing neutrino energy in NCQE is not reconstructed, we integrate Eq.~\eqref{eq:rate} over the entire beam spectrum, rather than considering a binned rate.

Although cancellations in $R^{\theta}$ are not exact, since the same target and detector
are utilized at every position  substantial components of these
uncertainties are expected to be correlated across positions. This is
a key advantage of combining the event rate ratio with detector motion to probe new physics.
The common normalization effects primarily shift all $R^\theta$, whereas
energy-dependent new physics effects can alter their relative
position dependence.

As a benchmark, we combine residual uncertainties from the flux,
interaction modeling, detector response, event selection  and
backgrounds into an effective fractional uncertainty on the ratio
$R^\theta$, taking $ \delta R_{\rm sys}=5\%,~10\%$.
This simplified treatment is intended to illustrate plausible future
IWCD performance. A complete uncertainty analysis requires a dedicated
detector-level study.

Although we focus on CCQE and NCQE interactions, pion-producing NC and
CC channels, including resonant production, could provide additional
observables sensitive to new physics effects beyond SM. In particular, single pion samples could provide a promising extension of the present
analysis. The ratio of NC$1\pi^0$ to CC$1\pi^0$ events
could benefit from strong correlations in the flux normalization,
$\pi^0$ reconstruction  and parts of the pion production model since
both channels are measured with the same target and detector. Their
resonant energy dependence may also provide a complementary response
to the changing off-axis spectra.
 Their inclusion requires
dedicated analyses that we leave for future work.

\section{Neutrino   Interactions}\label{sec:cross_sections}

\subsection{Quasi-Elastic Scattering Interactions}

To describe the CCQE and NCQE processes we follow 
Llewellyn-Smith formalism for scattering on a free nucleon at
rest as given in Ref.~\cite{LlewellynSmith:1971uhs,Formaggio:2012cpf}. We obtain the oxygen cross-section by summing free nucleon contributions and approximate small $Q^2$ nuclear suppression through the phenomenological prescription described below. The differential cross-section
per nucleon is
\begin{equation}\label{eq:dsdQ}
    \dfrac{d\sigma}{dQ^2} = \dfrac{G_F^2 M^2 |V_{ud}|^2}{8\pi E_\nu^2} \left[ A\pm \dfrac{(s-u)}{M^2}B + \dfrac{(s-u)^2}{M^4}C \right]~,
\end{equation}
where the plus (minus) sign applies to neutrino (antineutrino)
scattering, in the convention $F_A(0)=g_A$ discussed below. Here, $G_F \simeq 1.166 \times 10^{-5}$~GeV$^{-2}$ is the Fermi
constant, $M$ is the nucleon mass,   $Q^2=-q^2>0$ is the squared four-momentum transfer,  $(s-u)=4ME_\nu-Q^2-m_l^2$ 
with $m_l$ being the lepton mass. The $|V_{ud}|^2=0.94803$~\cite{ParticleDataGroup:2024cfk} Cabibbo-Kobayashi-Maskawa quark mixing matrix element is applied only to CCQE scattering, and set to unity for NCQE. For NCQE scattering, the outgoing neutrino mass is neglected  with
$m_l\simeq0$.

The functions $A$, $B$, and $C$ depend on the contributing   vector ($F_1$, $F_2$),
axial vector ($F_A$) and  pseudoscalar ($F_{PS}$) form factors of the nucleon as
\begin{align} \label{eq:FFs}
    A =&~ \dfrac{(m_l^2 + Q^2)}{M^2} \Big[~ (1+\tau) F_A^2 - (1-\tau) F_1^2   \nonumber \\
    &~ +~ \tau(1-\tau) F_2^2 + 4\tau F_1 F_2  ~ \nonumber\\
    &~-\frac{m_l^2}{4M^2}
\left\{
(F_1+F_2)^2
+(F_A+2F_{PS})^2
-4(1+\tau)F_{PS}^2
\right\} \Big]~ \nonumber\\
    B =&~ \dfrac{Q^2}{M^2} F_A (F_1 + F_2)~, \nonumber\\
    C =&~ \dfrac{1}{4} \left[ F_A^2 + F_1^2 + \tau F_2^2 \right]~,  
\end{align}
where $\tau = Q^2/4M^2$. The pseudoscalar contribution is proportional to the outgoing lepton
mass and therefore does not contribute to NCQE scattering with a
massless final state neutrino. 
The corresponding form factors are given by
\begin{align}\label{eq:pn_ff}
    F^i_1(Q^2) =&~ \frac{G^i_E(Q^2) + \tau G^i_M(Q^2)}{1+\tau}~, \nonumber\\
    F^i_2(Q^2) =&~ \frac{G^i_M(Q^2) - G^i_E(Q^2)}{1+\tau}~,  \nonumber\\
    F_A(Q^2) =&~ \frac{g_A}{(1+Q^2/M_A^2)^2}~, \nonumber \\
    F_{PS}(Q^2)=&~\frac{2M^2}{m_\pi^2+Q^2} F_A(Q^2)~,
\end{align}
where $i=\{p,n\}$, $m_\pi$ is the pion mass. We use
$g_A=1.2754$~\cite{ParticleDataGroup:2024cfk}  as well as  
$M_A=1.236~\mathrm{GeV}$, obtained from an average of lattice QCD
determinations of the axial form factor~\cite{Meyer:2026kdl}.

For the electromagnetic Sachs nucleon form factors we use the dipole 
parameterization for $G_E^p$, $G_M^p$  and $G_M^n$, together with
the Galster parameterization~\cite{Galster:1971kv} for $G_E^n$  
\begin{align}
    G_E^p(Q^2) &= \frac{1}{(1+Q^2/M_V^2)^2}~, \nonumber\\
    G_E^n(Q^2) &= \Big(\frac{-\mu_n \tau}{1 + 5.6\tau} \Big)\frac{1}{(1+Q^2/M_V^2)^2}~, \nonumber\\
    G_M^p(Q^2) &= \frac{\mu_p}{(1+Q^2/M_V^2)^2}~,  \nonumber\\
    G_M^n(Q^2) &= \frac{\mu_n}{(1+Q^2/M_V^2)^2}~,  
\end{align}
where $M_V\simeq 0.84$~GeV~\cite{Baker:1981tx, LlewellynSmith:1971uhs} is the vector mass and $\mu_p = 2.793$, $\mu_n = -1.913$ are the magnetic moments of the proton and neutron, respectively.

 \begin{figure}[t] 
    \centering
    \includegraphics[width=0.92\linewidth]{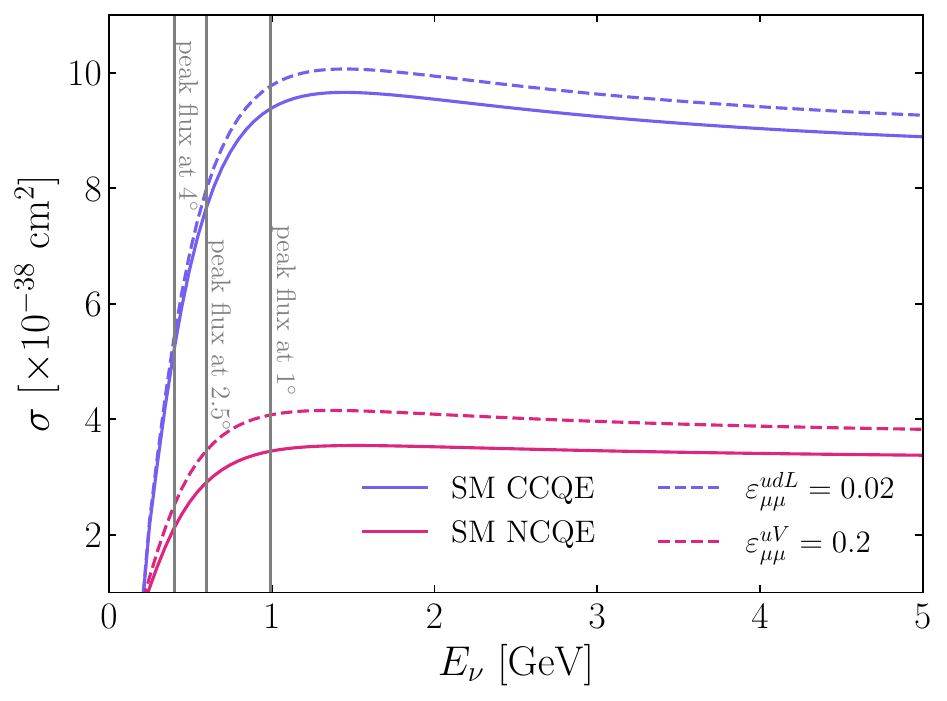}
    \caption{
   CCQE and NCQE cross-sections for $\nu_\mu$ interactions on a single $^{16}$O nucleus in water. The SM
predictions are compared with illustrative NSI benchmarks
$\varepsilon_{\mu\mu}^{udL}=0.02$ for CCQE and
$\varepsilon_{\mu\mu}^{uV}=0.2$ for NCQE. The
Pauli blocking cutoff is applied to both channels, while the CCQE
cross-section additionally includes the requirement that the outgoing
muon be above the Cherenkov threshold.
}
    \label{fig:cross_sec}
\end{figure}

For CCQE scattering  the   conserved vector current relation connects the weak vector form factors to the electromagnetic proton and neutron form factors ~\cite{LlewellynSmith:1971uhs,Formaggio:2012cpf}.
\begin{eqnarray}\label{eq:CC_ff}
F_{1,2}^{\rm CC}(Q^2)
=
F_{1,2}^{p}(Q^2)-F_{1,2}^{n}(Q^2). 
\end{eqnarray}
The corresponding CC axial and  pseudoscalar form factors are
\begin{equation}
F_A^{\rm CC}(Q^2)=F_A(Q^2)~~~,~~~ 
F_{PS}^{\rm CC}(Q^2)=F_{PS}(Q^2).
\end{equation}  

For NCQE scattering, the proton and neutron form factors are~\cite{Formaggio:2012cpf,Ilma:2024lkp}
\begin{align}
\label{eq:NC_ff}
F_m^{{\rm NC},p}
&=
\left(\frac{1}{2}-2\sin^2\theta_W\right)F_m^p
-\frac{1}{2}F_m^n
-\frac{1}{2}F_m^s,
\nonumber\\
F_m^{{\rm NC},n}
&=
\left(\frac{1}{2}-2\sin^2\theta_W\right)F_m^n
-\frac{1}{2}F_m^p
-\frac{1}{2}F_m^s, 
\end{align}
where $m = \{1,2\}$, while in the convention $F_A(0)=g_A$ the corresponding axial
form factors are
\begin{align}
F_A^{{\rm NC},p}
&=
\frac{1}{2}F_A-\frac{1}{2}F_A^s,
&
F_A^{{\rm NC},n}
&=
-\frac{1}{2}F_A-\frac{1}{2}F_A^s~,
\label{eq:NC_axial_ff}
\end{align}
where
$\sin^2\theta_W=0.22348$~\cite{ParticleDataGroup:2024cfk} is the weak mixing angle.
In the convention of Ref.~\cite{Abbaslu:2024jzo} instead the axial form factor is defined with the
opposite overall sign and this choice is compensated in the
vector-axial interference term, yielding the same physical cross
section.

The strange axial form factor is parameterized as
\begin{equation}
F_A^s(Q^2)
=
\frac{\Delta s}{\left(1+Q^2/M_A^2\right)^2},
\end{equation}
where $\Delta s$ denotes the strange quark contribution to the
nucleon spin. We consider $\Delta s=-0.08$ as a benchmark, based on the
combined analysis of Refs.~\cite{Pate:2006mv,Alberico:1998qw,HAPPEX:2004lbc}.
However, there remains significant uncertainty that includes compatibility with
$\Delta s=0$. Our results are not very sensitive to variations in $\Delta s$. 

\begin{figure}[t] 
    \centering
    \includegraphics[width=\linewidth]{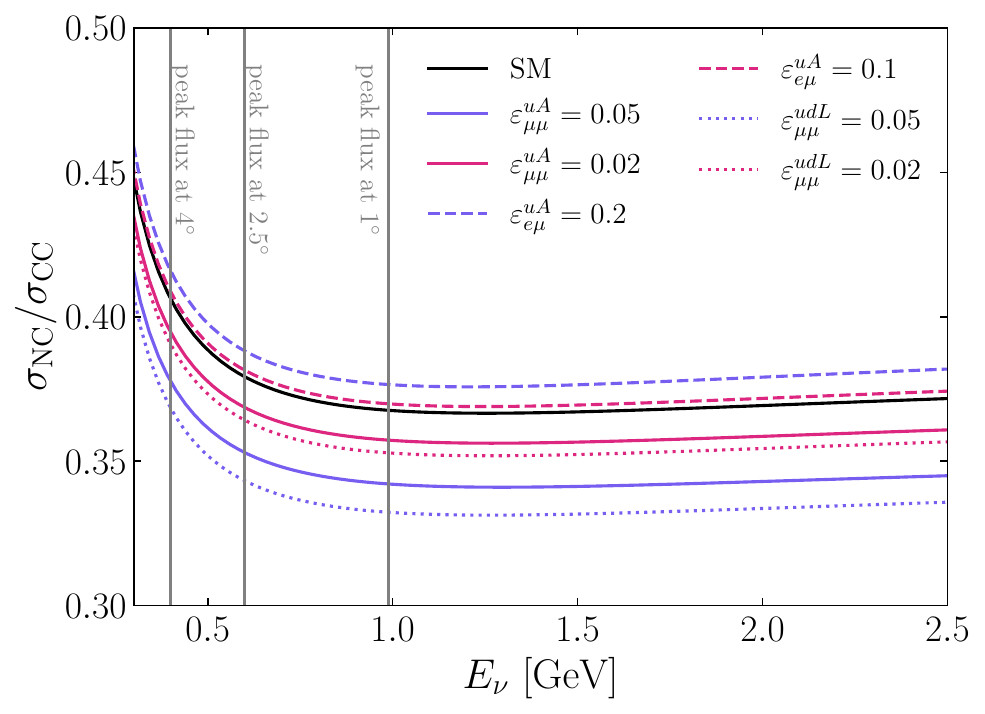}
    \caption{Ratio of the NCQE and CCQE cross-sections per oxygen nucleus as a
function of neutrino energy for representative NSI coefficients. For
NCQE scattering  the contributions from all eight protons and eight
neutrons in $^{16}\mathrm{O}$ are included. For CCQE scattering in
the neutrino-dominated FHC beam  only the eight neutrons contribute
through $\nu_\mu n\rightarrow\mu^-p$. The vertical lines indicate the
approximate peak energies of the fluxes at the three benchmark
off-axis positions. The same momentum transfer requirements as in
Fig.~\ref{fig:cross_sec} are applied.
}
    \label{fig:cross_ratio}
\end{figure}

The strange form factors are related to the strange
Sachs electromagnetic form factors $G_E^s$, $G_M^s$ through
\begin{align}
F_1^s(Q^2)
=&~
\frac{G_E^s(Q^2)+\tau G_M^s(Q^2)}{1+\tau},\notag\\
F_2^s(Q^2)
=&~
\frac{G_M^s(Q^2)-G_E^s(Q^2)}{1+\tau}.
\end{align}
We set $G_E^s(Q^2)$ to zero and use the benchmark dipole form $ 
G_M^s(Q^2)
=
\mu_s /(1 + Q^2/M_V^2)^2$ with strange magnetic
moment
$\mu_s\simeq-0.07$, 
consistent with lattice QCD determinations~\cite{Shanahan:2014tja}.

Within the incoherent free nucleon treatment adopted here the
cross-section for a nucleus containing $Z$ protons and $N$ neutrons
is obtained by summing the corresponding proton and neutron
contributions 
\begin{equation}
\label{eq:tot_cx}
\frac{d\sigma}{dQ^2}
=
Z\frac{d\sigma_p}{dQ^2}
+
N\frac{d\sigma_n}{dQ^2}.
\end{equation}
For water interactions as in the IWCD, one accounts for $^{16}\mathrm{O}$ with $Z=N=8$. Considering neutrino beam from T2HK FHC operation, the CCQE process on oxygen gives
\begin{equation}
\frac{d\sigma_{\rm CCQE}^{^{16}\mathrm{O}}}{dQ^2}
=
8\frac{d\sigma_{n}^{\rm CC}}{dQ^2}.
\end{equation}
For NCQE scattering, both proton and neutron contributions enter NCQE process on oxygen
\begin{equation}
\frac{d\sigma_{\rm NCQE}^{^{16}\mathrm{O}}}{dQ^2}
=
8\frac{d\sigma_{p}^{\rm NC}}{dQ^2}
+
8\frac{d\sigma_{n}^{\rm NC}}{dQ^2}.
\end{equation}
The two free hydrogen protons in each water molecule are not included in
the neutrino NCQE contribution.
In the case of antineutrino beam RHC operation, however, CCQE scattering also occurs on the two
free hydrogen protons through
$\bar\nu_\mu p\rightarrow\mu^+n$. In the present analysis we focus on
neutrino-dominated beam from FHC operation.
 
\subsection{Kinematics and Thresholds}

For the CCQE process $\nu_\mu n\rightarrow\mu^-p$ with the initial
neutron at rest the two-body scattering kinematic endpoints are
\begin{align}
\label{eq:Q2kin}
Q^{2,\mathrm{kin}}_{\min,\max}
&=
-m_\ell^2
+\frac{m_nE_\nu}{s}
\left[
s+m_\ell^2-m_p^2
\mp
\sqrt{\lambda(s,m_\ell^2,m_p^2)}
\right],
\end{align}
where $s=m_n^2+2m_nE_\nu$ 
and $\lambda(a,b,c)
=
a^2+b^2+c^2-2ab-2ac-2bc$  
is the K\"allen function. The upper sign in
Eq.~\eqref{eq:Q2kin} corresponding to the minus sign in front of the
square root gives the  minimum $Q^{2,\mathrm{kin}}_{\min}$, while the lower sign
gives the  maximum $Q^{2,\mathrm{kin}}_{\max}$. The energy threshold for CCQE  is
\begin{equation}
E_{\nu,\mathrm{th}}
=
\frac{(m_p+m_\ell)^2-m_n^2}{2m_n},
\end{equation}
and the cross-section is set to zero below it.

To approximately account for the suppression of low momentum transfer
interactions by Pauli blocking  we impose a phenomenological hard cutoff
\begin{equation}
\label{eq:Q2PB}
Q_{\rm PB}^2
=
(0.225~\mathrm{GeV})^2
\simeq 0.05~\mathrm{GeV}^2
\end{equation}
for both CCQE and NCQE scattering~\cite{SMITH1972605,Hayato:2009zz}.
This is a simplified treatment of the actual nuclear response. In a realistic
nucleus the nucleons have a momentum distribution due to
Fermi motion and Pauli blocking depends on both the initial state and
final state nucleon momenta. The resulting suppression is therefore
continuous rather than a sharp function of $Q^2$  and Fermi motion can
allow for some events below the nominal cutoff $Q_{\rm PB}^2$ considered here. More complete
calculations would include effects such as nucleon binding,  
correlations and final state interactions. Our hard cutoff thus represents a phenomenological benchmark rather than
an exact threshold.

Considering detection, for the event level CCQE data sample we additionally require the outgoing
muon to be above the Cherenkov threshold in water. We take $ 
E_{\mu,\mathrm{Ch}}\simeq160~\mathrm{MeV}$,
corresponding to a muon momentum of approximately
$120~\mathrm{MeV}$~\cite{ParticleDataGroup:2024cfk}. For a free
neutron initially at rest, energy conservation relates this
requirement to the upper bound
\begin{equation}
\label{eq:Q2Ch}
Q_{\mathrm{Ch}}^2(E_\nu)
=
2m_n\left(E_\nu-E_{\mu,\mathrm{Ch}}\right)
+m_n^2-m_p^2 .
\end{equation}
Thus, we employ the integration range for the detectable CCQE events as
\begin{align}
\label{eq:Q2limitsCC}
Q_{\mathrm{low,CC}}^2
&=
\max\left[
Q^{2,\mathrm{kin}}_{\min},
Q_{\rm PB}^2
\right],
\nonumber\\
Q_{\mathrm{high,CC}}^2
&=
\min\left[
Q^{2,\mathrm{kin}}_{\max},
Q_{\mathrm{Ch}}^2
\right].
\end{align}
The accepted contribution is set to zero whenever
$Q_{\mathrm{high,CC}}^2\leq Q_{\mathrm{low,CC}}^2$.

For  NCQE scattering $\nu N\rightarrow\nu N$, with
$N=\{p,n\}$ initially at rest and the neutrino mass neglected, the
kinematic endpoints are
\begin{equation}
Q^{2,\mathrm{kin}}_{\min}=0~,
\qquad
Q^{2,\mathrm{kin}}_{\max,N}
=
\frac{4m_NE_\nu^2}{m_N+2E_\nu}.
\end{equation}
Since we are considering detection of NCQE through de-excitation emission of visible $\gamma$ from the excited nucleus after the interaction we do not impose any additional Cherenkov threshold requirements\footnote{This does not imply the absence of neutrino-proton NC elastic scattering.}. Therefore, we consider NCQE integration range  
\begin{equation}
\label{eq:Q2limitsNC}
Q_{\mathrm{low,NC}}^2=Q_{\rm PB}^2~,
\qquad
Q_{\mathrm{high,NC}}^2
=
Q^{2,\mathrm{kin}}_{\max,N}~,
\end{equation}
with the proton and neutron masses used for their respective
contributions. The NCQE contribution is set to zero whenever $Q^{2,\mathrm{kin}}_{\max,N} \leq Q_{\rm PB}^2$.

\section{Neutrino Non-Standard Interactions}\label{sec:NSI}

\subsection{Effective Operators and Form Factors}

NSI can modify
neutrino interaction cross-sections and alter the
event rates at IWCD. NC and CC NSI can affect both the vector and axial-vector contributions. 
Here, we treat the NSI contributions as phenomenological low energy parameters and remain agnostic about their specific ultraviolet origin. In a specific gauge-invariant completion, several such NSI coefficients can be correlated and additional constraints or interactions can thus arise.

We focus on the heavy mediator limit in which the mediator mass satisfies $m_{\rm med}^2 \gg Q^2$. The new interactions can then be described
by effective four-fermion operators and the effective NC NSI Lagrangian contributions are given by~\cite{Proceedings:2019qno}
\begin{equation}
\label{eq:ncnsilagrangian}
\mathcal{L}_{\rm NSI}^{\rm NC}
=
-2\sqrt{2} G_F
\varepsilon_{\alpha\beta}^{fP}
\left(\bar{\nu}_\alpha\gamma^\mu P_L\nu_\beta\right)
\left(\bar{f}\gamma_\mu P f\right),
\end{equation}
where $\alpha,\beta=\{e,\mu,\tau\}$ denote neutrino flavors, $f$ denotes
a SM fermion, $P=\{L,R\}$ where $P_L=(1-\gamma^5)/2$ and $P_R=(1+\gamma^5)/2$ are the chiral projection operators, and $\varepsilon_{\alpha\beta}^{fP}$
parameterizes the strength of the interaction relative to the Fermi
constant. We focus on $f$ being the first  generation
quarks $u$ and $d$.  

The operators in
Eq.~\eqref{eq:ncnsilagrangian} can induce flavor-changing NC (FCNC)
interactions for   non-diagonal $\varepsilon_{\alpha\beta}^{fP}$ with $\alpha\neq\beta$,   which can affect neutrino propagation in matter and
therefore neutrino oscillation observables
~\cite{Farzan:2017xzy,Esteban:2018ppq,Chatterjee:2020kkm,
Chatterjee:2024kbn,Esteban:2020cvm,Blennow:2016etl}.
Direct scattering measurements  provide complementary sensitivity,
particularly to flavor-diagonal NSI and to axial combinations that do
not contribute to the coherent matter potential relevant for oscillation matter effects in neutrino propagation. Here we primarily focus on
these flavor-conserving interactions, while also considering selected
off-diagonal couplings below.

The CC NSI are described by the effective Lagrangian
\begin{equation}
\label{eq:cc_nsi_lagrangian}
\mathcal{L}_{\rm NSI}^{\rm CC}
=
-2\sqrt{2} G_F V_{ud}
\varepsilon_{\alpha\beta}^{udP}
\left(\bar{\ell}_{\alpha}\gamma^\mu P_L\nu_{\beta}\right)
\left(\bar{u}\gamma_\mu P d\right)
+{\rm h.c.},
\end{equation}
where $\ell_\alpha$ is the charged lepton of flavor $\alpha$, and
$\varepsilon_{\alpha\beta}^{udP}$ parameterizes the interaction
relative to the SM CC amplitude. With the convention of
Eq.~\eqref{eq:cc_nsi_lagrangian}  the second flavor index labels the
incoming neutrino, while the first labels the associated outgoing
charged lepton.

Unlike NC NSI, CC NSI do not generate a coherent NC
matter potential during neutrino propagation. They can nevertheless
affect oscillation experiment measurements by modifying neutrino production at
the source and CC interaction detection at the detector. Direct
scattering and decay measurements therefore provide  
complementary probes of these interactions
~\cite{Farzan:2017xzy,Proceedings:2019qno}.

For both NC and CC interactions  the chiral quark couplings can be
expressed in terms of vector and axial-vector combinations 
\begin{align}
\label{eq:VA_nsi}
\varepsilon_{\alpha\beta}^{fV}
&=
\varepsilon_{\alpha\beta}^{fR}
+
\varepsilon_{\alpha\beta}^{fL},
\\
\varepsilon_{\alpha\beta}^{fA}
&=
\varepsilon_{\alpha\beta}^{fR}
-
\varepsilon_{\alpha\beta}^{fL}, \notag
\end{align}
where $f=\{u,d\}$ for NC NSI and $f=(ud)$ for CC NSI. Thus, the quark current can be expressed in terms of vector and
axial-vector components. Vector NSI modify the vector nucleon form
factors, while axial-vector NSI modify the corresponding axial form
factors.

Non-standard CC operators with Lorentz structures different from the
SM left-handed $(V-A)$ current are strongly constrained by precision
measurements of meson, nuclear  and charged-lepton decays
~\cite{Biggio:2009nt,Proceedings:2019qno}. We therefore focus in
present analysis on a left-handed CC interaction with the same
Lorentz structure as the SM amplitude. Then, the relevant
form factors are rescaled by $(1+\varepsilon_{\mu\mu}^{udL})$, and the
cross-section is proportional to
$(1+\varepsilon_{\mu\mu}^{udL})^2$. The leading deviation is therefore
linear in $\varepsilon_{\mu\mu}^{udL}$ for
$|\varepsilon_{\mu\mu}^{udL}|\ll1$, with additional contributions at larger couplings. 

For NC NSI, independent vector and axial-vector interactions modify in different ways
the functions $A$, $B$ and $C$ in Eq.~\eqref{eq:FFs}.
In particular, the vector-axial interference term $B$ produces an
energy-dependent response that can distinguish vector and axial
contributions when measurements at several off-axis spectra are
combined as we consider. This spectral dependence provides additional motivation
for exploiting the mobility of the IWCD to probe new physics effects.

In the main analysis we express the NC NSI coefficients in the
vector and axial vector basis and vary one quark coupling at a time,
except in the two parameter vector-axial  scan discussed below.
From Eq.~\eqref{eq:ncnsilagrangian}  an incident $\nu_\mu$ beam,
as relevant for the IWCD with neutrino beam configuration, probes coefficients
$\varepsilon_{\alpha\mu}^{qX}$, where
$q=\{u,d\}$, $X=\{V,A\}$ and flavor $\alpha$.
The flavor-diagonal interaction is described by
$\varepsilon_{\mu\mu}^{qX}$, while
$\varepsilon_{e\mu}^{qX}$ and
$\varepsilon_{\tau\mu}^{qX}$ induce FCNC
scattering. Since the flavor of the outgoing neutrino is not observed and
neutrino masses are neglected the scattering kinematics are identical
for $\nu_e$ and $\nu_\tau$ final states. Thus, we discuss the sensitivity for
off-diagonal couplings in terms of
$\varepsilon_{e \mu}^{qX}$, setting
$\varepsilon_{\tau\mu}^{qX}=0$. Varying one coupling at a time, the same analysis applies to
$\varepsilon_{\tau\mu}^{qX}$.
 The $u$-quark and $d$-quark coefficients can alternatively be reorganized
into isoscalar and isovector combinations, which we discuss in
App.~\ref{app:isospin}.

For CC NSI, we restrict the analysis to the flavor-diagonal 
left-handed coefficient $\varepsilon_{\mu\mu}^{udL}$. We do not consider CC
flavor-changing interactions that are generally subject to stringent constraints from charged lepton flavor
violation and precision decay measurements, although these constraints
can be model dependent~\cite{Biggio:2009nt}. Here, coefficients labeled by $V$ or $A$ refer to NC NSI, while the
CC interaction is denoted   by
$\varepsilon_{\mu\mu}^{udL}$.

Following Refs.~\cite{Ilma:2024lkp,Abbaslu:2024jzo}, the modified NSI NC nucleon form factors can be
written as
\begin{align}\label{eq:NSI_ff}
F_m^{{\rm NC}\varepsilon, (p/n)}
&= \left(\frac{\delta_{\alpha\beta}}{2}
    - 2\delta_{\alpha\beta}\sin^2\theta_W
    + 2\varepsilon_{\alpha\beta}^{uV}
    + \varepsilon_{\alpha\beta}^{dV}\right)
    F_m^{(p/n)}
    \notag\\
&\quad
 + \left(-\frac{\delta_{\alpha\beta}}{2}
    + \varepsilon_{\alpha\beta}^{uV}
    + 2\varepsilon_{\alpha\beta}^{dV}\right)
    F_m^{(n/p)}
 \notag\\
&\quad
 + \left(-\frac{\delta_{\alpha\beta}}{2}
    + \varepsilon_{\alpha\beta}^{uV}
    + \varepsilon_{\alpha\beta}^{dV}\right)
    F_m^s~, \notag
\\[1ex]
F_A^{{\rm NC}\varepsilon, (p/n)}  
&= \pm\left(\frac{\delta_{\alpha\beta}}{2}
    - \frac{\varepsilon_{\alpha\beta}^{uA}
    - \varepsilon_{\alpha\beta}^{dA}}{2}\right)
    F_A
  \\
&\quad
 - \frac12
    \left(\varepsilon_{\alpha\beta}^{uA}
    + \varepsilon_{\alpha\beta}^{dA}\right)
    F_A^{(8)}
 \notag\\
&\quad
 - \left(\frac{\delta_{\alpha\beta}}{2}
    + \varepsilon_{\alpha\beta}^{uA}
    + \varepsilon_{\alpha\beta}^{dA}\right)
    F_A^s~,  \notag
\end{align}
where $\delta_{\alpha\beta}$ denotes the Kronecker delta function, $m=\{1,2\}$, and the upper (lower) sign corresponds to a proton
(neutron). We neglect NSI involving strange quarks, with
$\varepsilon_{\alpha\beta}^{sV}
=\varepsilon_{\alpha\beta}^{sA}=0$, while retaining the SM strange
nucleon form factors $F_i^s$ and $F_A^s$.

The octet axial form factor $F_A^{(8)}$ describes the nucleon matrix element of the flavor-octet axial current\footnote{Following normalization of Ref.~\cite{Alexandrou:2021wzv}  we define $F_A^{(8)} = F_A^u + F_A^d - 2 F_A^s$. In this convention  the coefficient of the isoscalar axial NSI combination in Eq.~\eqref{eq:NSI_ff} is 1/2. The coefficient of 3/2 used in Ref.~\cite{Abbaslu:2024jzo} corresponds to an additional factor 1/3 in the current normalization and is equivalent when $g_A^{(8)}$ is rescaled appropriately.}.
It enters the modified NC form factor in
Eq.~\eqref{eq:NSI_ff} through the isoscalar axial NSI combination
$\varepsilon_{\alpha\beta}^{uA}
+\varepsilon_{\alpha\beta}^{dA}$. We again use the dipole
parameterization
\begin{eqnarray}
    F_A^{(8)}(Q^2) = \frac{g_A^{(8)}}{\left(1+Q^2/(M_A^{(8)})^2\right)^2}~,
\end{eqnarray}
with $g_A^{(8)}=0.53$ and $M_A^{(8)}=1.154$~GeV~\cite{Abbaslu:2024jzo, Alexandrou:2021wzv}.   

Within the effective operator framework that we consider NC NSI do
not modify the charged meson decays that are relevant for producing  neutrino beams at
tree level and therefore enter directly through the NCQE scattering
amplitude at detector. For diagonal NC NSI, the interference with the SM also
provides sensitivity to the sign of the coupling. 

For CC NSI we restrict the analysis to the flavor-diagonal
left-handed coefficient $\varepsilon_{\mu\mu}^{udL}$, whose Lorentz
structure is identical to that of the SM charged current. The  
CCQE amplitude is then rescaled as
\begin{align} \label{eq:CC_ff_nsi}
    F_Y^{{\rm CC}\varepsilon}(Q^2) =&~ \left(1 + \varepsilon_{\mu\mu}^{udL}\right) F^{\rm CC}_Y (Q^2)~,  
\end{align}
where $Y = \{1,2,A,PS \}$.
Consequently, the SM cross-section of Eq.~\eqref{eq:dsdQ} is rescaled 
\begin{equation}
\label{eq:CCsigmansi}
\frac{d\sigma^{\mathrm{CC}\varepsilon}}{dQ^2}
=
(1+\varepsilon_{\mu\mu}^{udL})^2
\frac{d\sigma^{\mathrm{CC}}}{dQ^2}.
\end{equation}
We take the NSI coefficients to be real in the numerical analysis.

The diagonal left-handed CC contributions can also modify the
meson $\pi^+$ and $K^+$ decays that produce the dominant
$\nu_\mu$  beam component. For the same
$\varepsilon_{\mu\mu}^{udL}$ operator, this produces a common
multiplicative change in the incident  $\nu_\mu$  flux. Such a common
source factor cancels between the NCQE numerator and CCQE denominator
of $R^\theta$. This further motivates our implementation  of
$\varepsilon_{\mu\mu}^{udL}$ through the CCQE detection cross-section in the ratio observable established in
Eq.~\eqref{eq:rate}.

We do not consider off-diagonal CC coefficients in the present
analysis. For an incident $\nu_\mu$, the coefficient
$\varepsilon_{\tau\mu}^{udL}$ produces an outgoing $\tau$, whose
production threshold is approximately $\sim 3.5 ~\mathrm{GeV}$, where
the J-PARC off-axis beam flux is significantly suppressed. On the other hand, the coefficient
$\varepsilon_{e\mu}^{udL}$ instead produces an outgoing electron.
Such flavor-changing CC interactions are often strongly constrained
in gauge invariant ultraviolet completions by charged lepton flavor
observables, although this depends on the model~\cite{MEG:2016leq,Gavela:2008ra}. We leave detailed treatment of these effects for future work.  

\begin{figure}[t] 
    \centering
    \includegraphics[width=\linewidth]{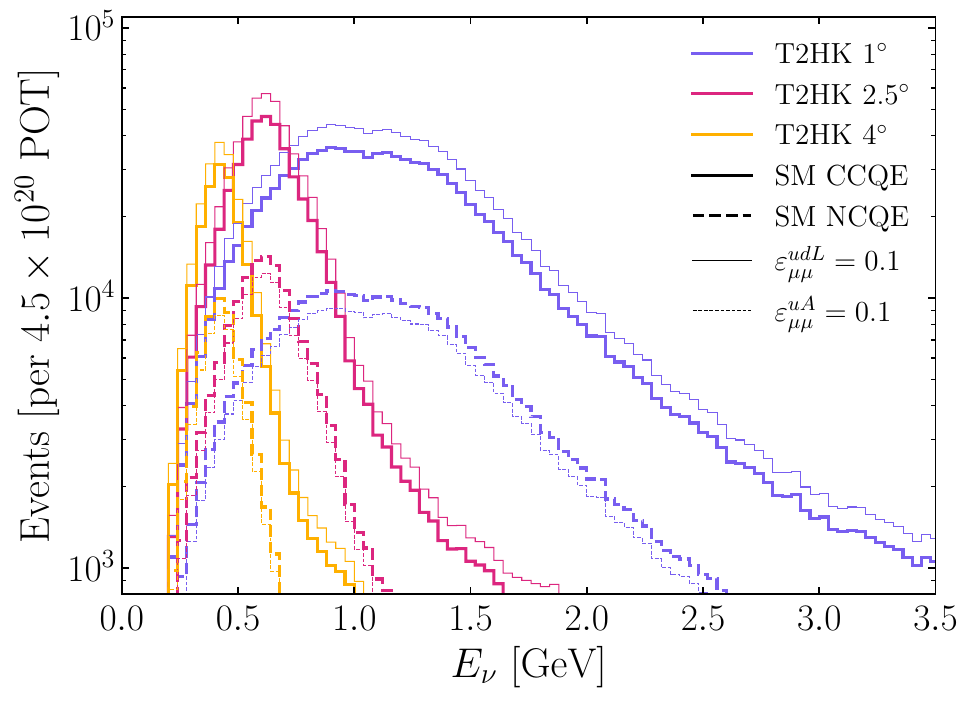}
    \caption{
Illustrative CCQE and NCQE event spectra at the $1^\circ$,
$2.5^\circ$, and $4^\circ$ off-axis angle IWCD positions for the T2HK neutrino beam. Each spectrum is shown
for an exposure of $4.5\times10^{20}$ POT to facilitate comparison
among the incident spectral shapes. The spectral flux shapes are obtained from Ref.~\cite{Naseby:2021olw} while the normalization is determined by matching to the total true CCQE event counts as in Tab. III of Ref.~\cite{nuPRISM:2014mzw}, for   each corresponding off-axis interval. 
The NCQE rates use the same flux normalization
together with an energy independent benchmark detection efficiency of
$80\%$. Thin curves illustrate the effects of
$\varepsilon_{\mu\mu}^{uA}=0.1$ in NCQE scattering and
$\varepsilon_{\mu\mu}^{udL}=0.1$ in CCQE scattering. In the
sensitivity analysis, the total exposure is instead held fixed and
divided equally among the detector positions included in each
configuration.    
  }
    \label{fig:event_rates}
\end{figure}
 
\subsection{Benchmark Effects at IWCD}
 
We first illustrate how the NSI operators modify the relevant
interaction observables. The CCQE and NCQE cross sections are shown\footnote{We note that our  simplified free nucleon treatment yields an absolute NCQE cross-section somewhat larger than the nominal NEUT predictions of the T2K analysis~\cite{T2K:2019zqh}.  
However, the comparison is not direct as  nuclear binding, Fermi motion, a spectral function description of the initial state, and intranuclear interactions are used in NEUT, whereas we approximate oxygen as an incoherent sum over nucleons with an effective $Q^2$ cutoff. We therefore use the resulting rates as benchmarks for the relative NSI response across off-axis spectra. Detailed quantitative predictions for these event samples require dedicated simulations and are left for future work.}
in Fig.~\ref{fig:cross_sec}, while their ratio is shown in
Fig.~\ref{fig:cross_ratio}. Only the calculated interaction
cross-sections are considered without detector-related efficiencies, which become relevant when constructing the observable event
rates discussed below.

For flavor-diagonal NC NSI  the SM and NSI amplitudes interfere, thus
the leading cross-section modification is linear in
$\varepsilon_{\mu\mu}^{qX}$ for sufficiently small couplings.
Vector and axial interactions modify the functions $A$, $B$  and
$C$ in Eq.~\eqref{eq:dsdQ} and Eq.~\eqref{eq:FFs} differently. In
particular, the vector-axial   interference contribution $B$ carries
a different energy and momentum-transfer dependence from the purely
vector and axial-vector terms. Thus, combining measurements at multiple off-axis detector positions can
distinguish an NSI-induced spectral distortion from just a
common normalization shift and can separate vector and axial vector
contributions that can remain degenerate for a single incident spectrum.
This contrasts with the left-handed CC NSI, which uniformly
rescales the SM CCQE amplitude.

For off-diagonal NC NSI  the final-state neutrino flavor differs from
the incoming flavor.  Since these
final state neutrino flavors are orthogonal the SM and off-diagonal
NSI amplitudes do not interfere. The off-diagonal contribution
therefore enters the cross section quadratically in the NSI
coefficient. In convention of Eq.~\eqref{eq:ncnsilagrangian}, the second
flavor index denotes the incoming neutrino and the first denotes the
outgoing neutrino. An incident $\nu_\mu$ beam then probes
$\varepsilon_{\alpha\mu}^{qX}$, with $\alpha=\{e,\tau\}$. Since the
outgoing neutrino flavor is not measured, the total observable cross section
is the incoherent sum
\begin{equation}
\label{eq:offdiag_sigma}
\sigma_{\rm tot}
=
\sigma_{\rm SM}(\nu_\mu\to\nu_\mu)
+
\sum_{\alpha\neq\mu}
\sigma_{\rm NSI}(\nu_\mu\to\nu_\alpha).
\end{equation}
Varying in our analysis   one NSI coupling at a time, we consider effects of
$\varepsilon_{e\mu}^{qX}$ and set
$\varepsilon_{\tau\mu}^{qX}=0$.  In case both of these coefficients are nonzero  their contributions add without
interference as they produce different final-state flavors.

In Fig.~\ref{fig:event_rates} we illustrate the
corresponding CCQE and NCQE event spectra at the three benchmark
off-axis detector positions. For this
illustrative comparison, the same exposure is assigned to each angle.
The CCQE event distribution is determined from the un-normalized fluxes reported in~\cite{Naseby:2021olw} while the normalization is matched to the prediction of total CCQE events from
Ref.~\cite{nuPRISM:2014mzw}.  The NCQE event rate is obtained using the
same flux normalization together with the constant benchmark
efficiency $\eta_{\rm NCQE}=0.8$, following Ref.~\cite{T2K:2019zqh}. This efficiency is applied only to
the event level quantities and not to the cross sections in
Fig.~\ref{fig:cross_sec} and Fig.~\ref{fig:cross_ratio}. The
sensitivity analysis below instead keeps the total exposure fixed and
divides it equally among the detector positions included in each
configuration.

\begin{figure*}[t]
    \centering
    \begin{minipage}{0.48\textwidth}
        \centering
        \includegraphics[width=0.95\linewidth]{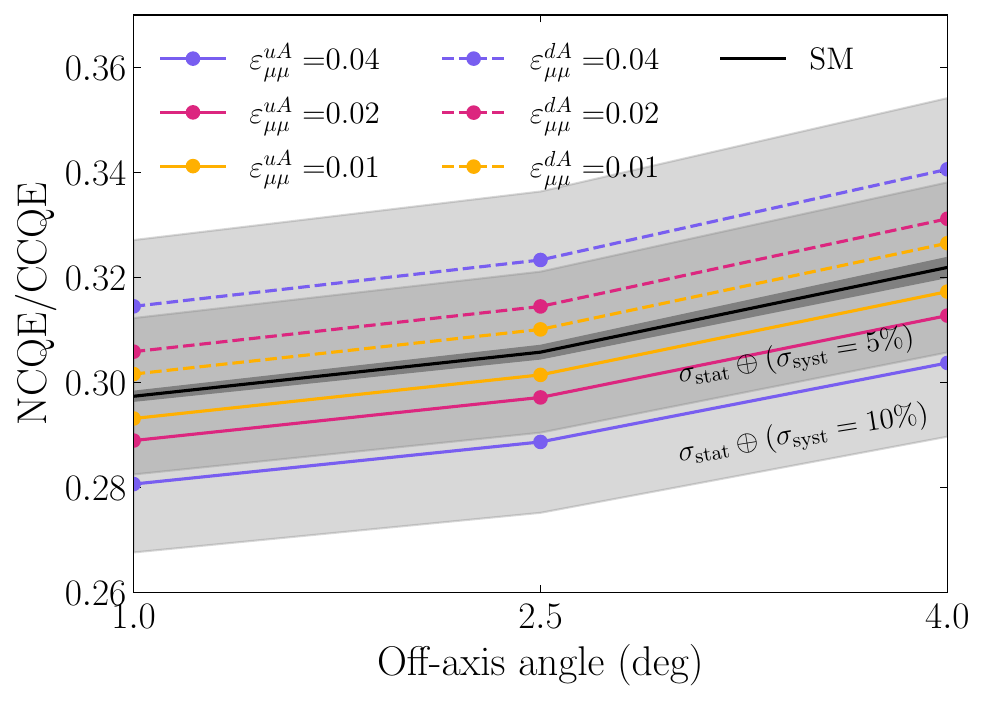}\\
        \includegraphics[width=0.95\linewidth]{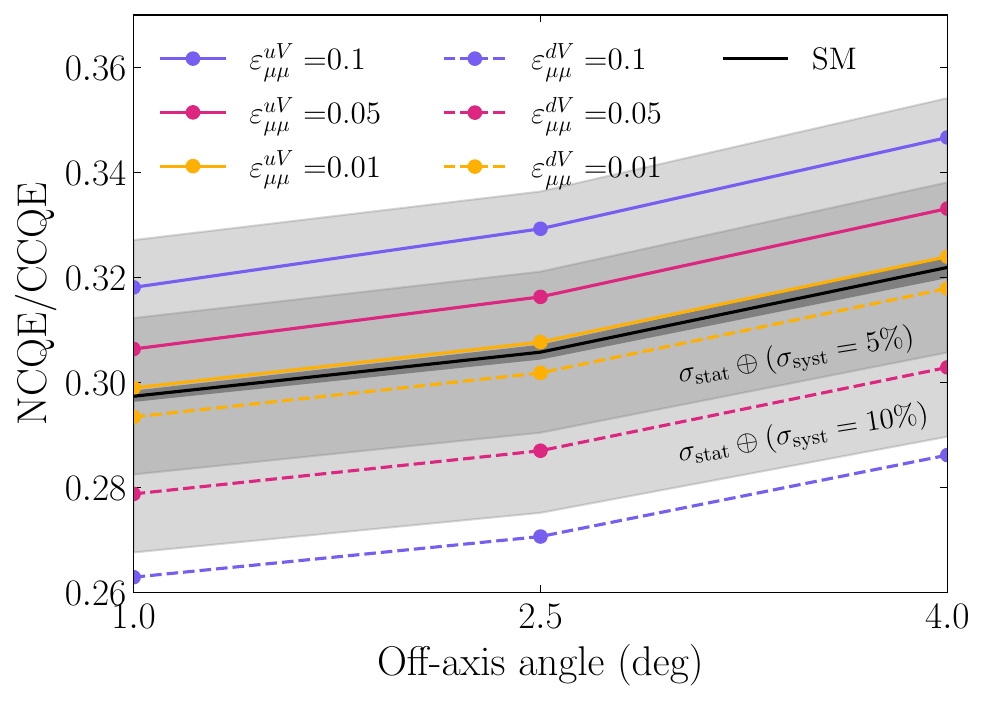}
    \end{minipage}
    \hfill
    \begin{minipage}{0.48\textwidth}
        \centering
        \includegraphics[width=0.95\linewidth]{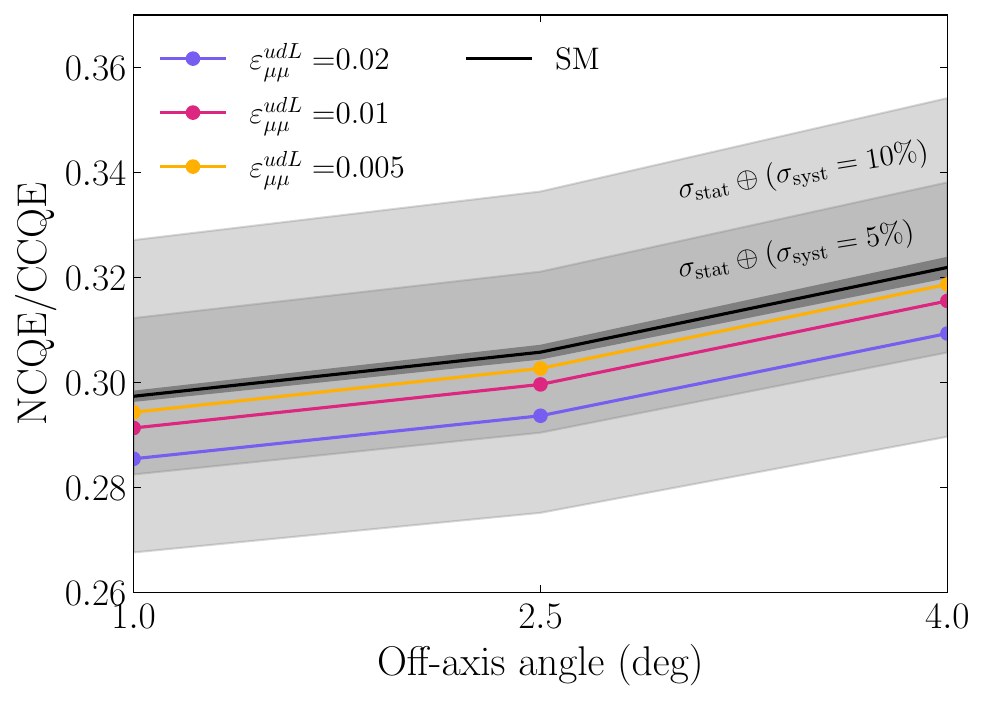}\\
        \includegraphics[width=0.95\linewidth]{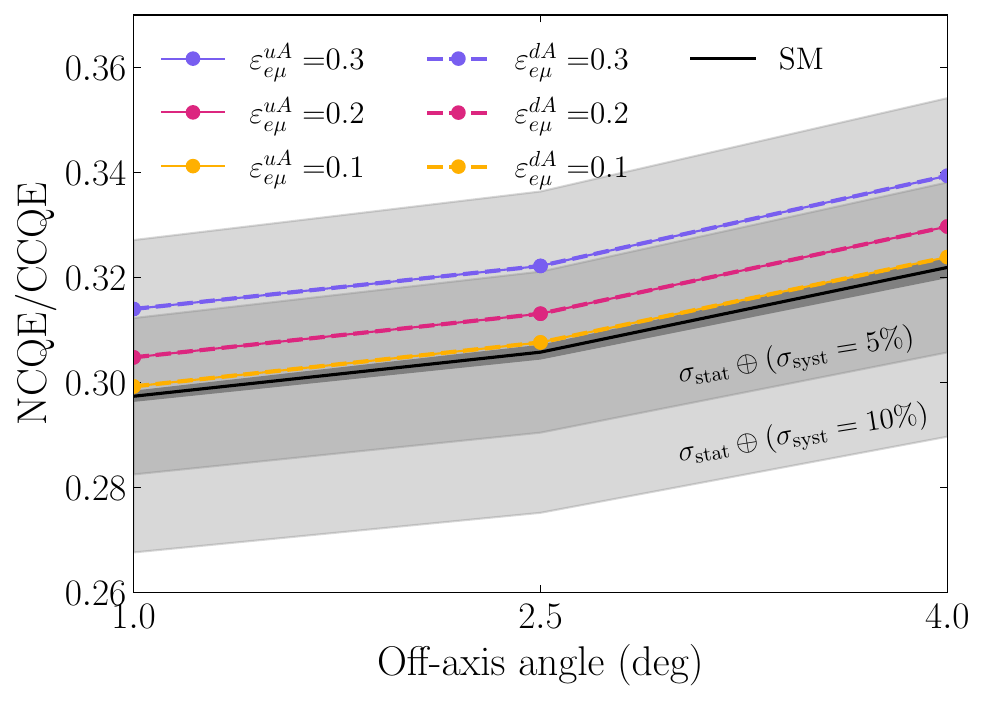}
    \end{minipage}
    \caption{
    NCQE to CCQE event rate ratios 
$R^\theta=N^\theta_{\rm NCQE}/N^\theta_{\rm CCQE}$  at the
$1^\circ$, $2.5^\circ$  and $4^\circ$ off-axis angle IWCD positions. The left
panels indicate flavor-diagonal axial (top) and vector (bottom) NC NSI,
while the right panels show the flavor-diagonal left-handed CC
coefficient $\varepsilon_{\mu\mu}^{udL}$ (top) and an off-diagonal
axial NC coefficient producing a $\nu_e$ final state (bottom). In the
NC panels, solid and dashed curves denote couplings to $u$ and $d$
quarks, respectively. 
Systematic and statistical uncertainties are included in the sensitivity analysis and are shown in shaded bands with the innermost dark gray band denoting purely statistical Poisson uncertainties and the two outer bands showing statistical uncertainties combined in quadrature with  $5\%$ and $10\%$ fractional systematic uncertainties, labeled by $\sigma_{\rm stat} \oplus \sigma_{\rm sys}$. Vector and axial NC interactions modify the dependence of
$R^\theta$ on the off-axis position, whereas the left-handed CC
interaction predominantly rescales the denominator. Off-diagonal
coefficients such as $\varepsilon_{e\mu}^{qA}$ enter quadratically
because their amplitudes do not interfere with the SM amplitude.   
}
    \label{fig:event_ratios}
\end{figure*}

Integrating the spectra over neutrino energy gives the resulting NCQE/CCQE event rate
ratios $R^\theta$ at the three off-axis angle IWCD positions in Fig.~\ref{fig:event_ratios},  including Poisson statistical
uncertainties.  
 Both vector
and axial NC NSI can modify the dependence of $R^\theta$ on the
off-axis detector position, whereas the left-handed CC interaction primarily
changes its overall normalization. The off-diagonal NC coefficients enter
quadratically, as the corresponding final state neutrino has no
SM amplitude and hence there is no interference as discussed above. 
The variation of $R^\theta$ across detector positions provides the
spectral information used in the statistical analysis below.

 \begin{figure*}[t] 
    \centering
    \includegraphics[width=0.43\textwidth]{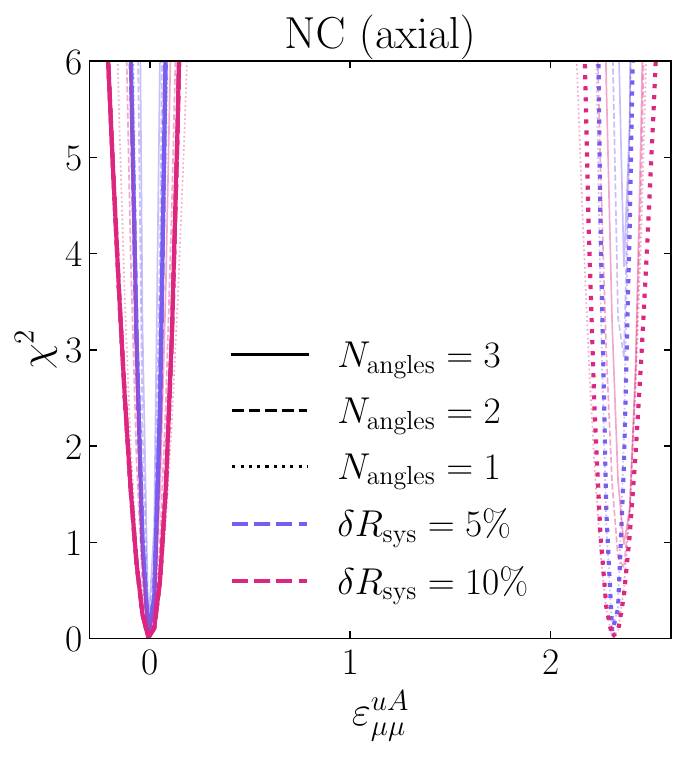}
    \includegraphics[width=0.43\textwidth]{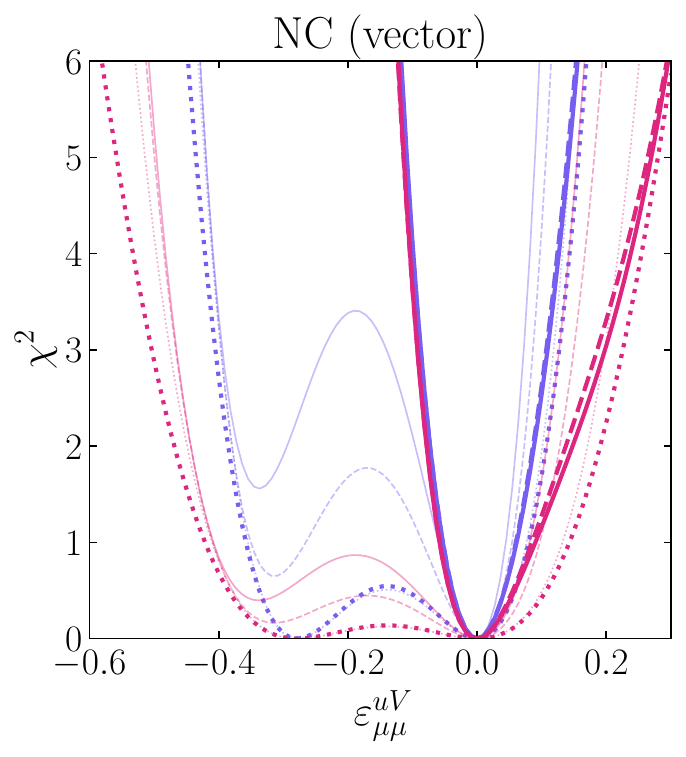}
    \includegraphics[width=0.43\textwidth]{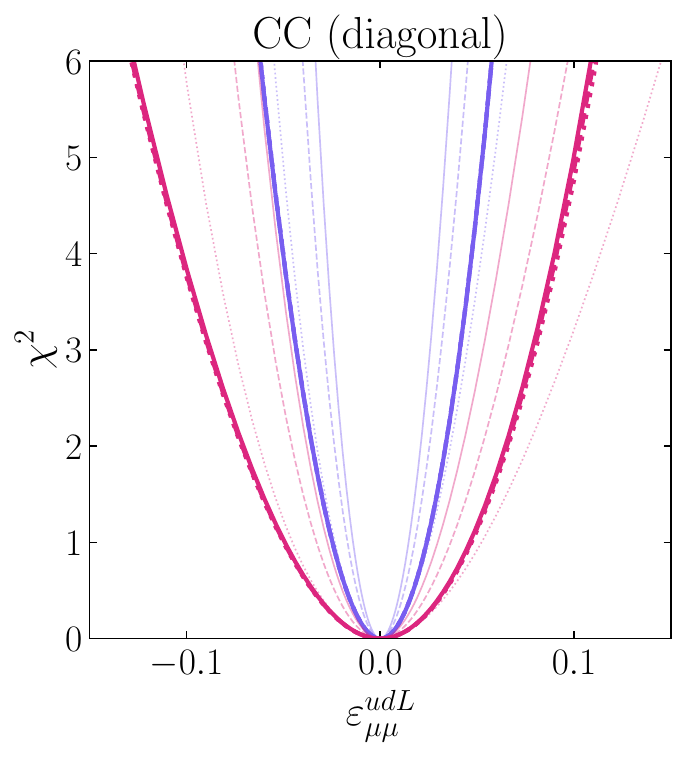}
    \includegraphics[width=0.43\textwidth]{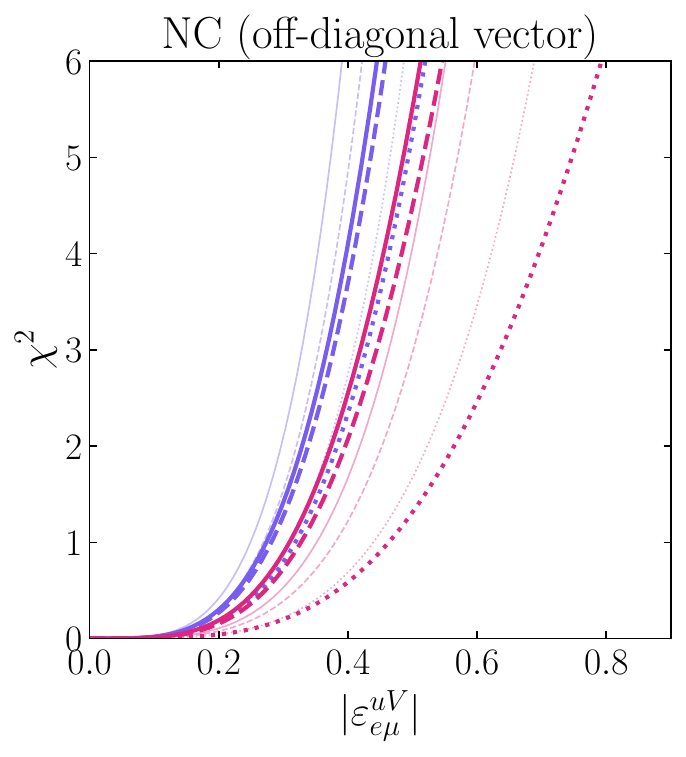}
    \caption{ 
Single parameter $\chi^2$ fits for flavor-diagonal axial NC NSI
(upper left), flavor-diagonal vector NC NSI (upper right), the
left-handed CC coefficient $\varepsilon_{\mu\mu}^{udL}$ (lower left) 
and the off-diagonal vector NC coefficient
$\varepsilon_{e\mu}^{uV}$ (lower right). Thin curves depict the
simplified independent position treatment with
Eq.~\eqref{eq:chi2base}, while thick curves show the
correlated normalization treatment of Eq.~\eqref{eq:pulled}. Line
styles indicate whether one, two, or three off-axis angle detector positions are
considered, with the same total exposure divided equally among the
selected positions. The two colors correspond to
$\delta R_{\rm sys}=5\%$ (blue) and $10\%$ (magenta). The off-diagonal result is shown
as a function of $|\varepsilon_{e\mu}^{uV}|$ since the event rate
depends quadratically on this coefficient. 
}
    \label{fig:chi2}
\end{figure*}

\section{Statistical Analysis and Sensitivity}\label{sec:results}

We evaluate the projected sensitivity to NSI considering movability of IWCD and that the measured 
NCQE to CCQE event ratios are in agreement with their SM expectations. At each
off-axis position angle $\theta = \{1^\circ,2.5^\circ,4^\circ\}$, we consider the ratio $R^{\theta}$. 
Assuming independent Poisson fluctuations in the selected NCQE and
CCQE event samples, the fractional statistical uncertainty is
\begin{align}
\label{eq:ratio_stat}
\left(\delta R_{{\rm stat} }^{\theta}\right)^2
=
\left(
\frac{\Delta R_{{\rm stat}}^{\theta}}{R^{\theta, \rm SM}}
\right)^2 =
\frac{1}{N_{{\rm NCQE}}^{\theta,\rm SM}}
+
\frac{1}{N_{{\rm CCQE}}^{\theta,\rm SM}} .
\end{align}
For an exposure of $4.5 \times 10^{20}$~POT at each individual detector position, we find the fractional statistical uncertainties  of approximately 0.3$\%$, 0.40$\%$  and 0.6$\%$ at 1$^\circ$, 2.5$^\circ$  and 4$^\circ$ off-axis positions, respectively. 
For multi-position analyses, when a fixed total exposure is divided among $n$ detector off-axis positions, the event counts at each position are reduced by $1/n$, and the corresponding statistical uncertainties are recomputed for that configuration.    For comparisons involving one ($\theta = 1^\circ$), two ($\theta = 1^\circ, 2.5^\circ$)  or three detector positions, the
total exposure is held fixed and divided equally among the positions
included in each configuration. The event numbers and statistical
uncertainties are   separately considered for each configuration.

We first consider a treatment in which the residual
systematic uncertainty is taken to be independent at each position 
\begin{equation}
\label{eq:chi2base}
\chi^2_{\rm base}(\varepsilon)
=
\sum_\theta
\frac{
\left[
R^{\theta,\rm SM}
-
R^{\theta,\rm NSI}(\varepsilon)
\right]^2
}{
\left(\delta R_{{\rm stat} }^{\theta}R^{\theta,\rm SM}\right)^2
+
\left(\delta R_{\rm sys}R^{\theta,\rm SM}\right)^2
}.
\end{equation}
We consider benchmark fractional systematic uncertainties
$\delta R_{\rm sys}=5\%$ and $10\%$.

We also consider that the dominant residual uncertainties are correlated
among detector positions since features such as common target, detector response  and
reconstruction procedure are expected to be the same at each location. We therefore also introduce
a common normalization nuisance parameter $f$ and define pulled $\chi^2$
\begin{equation}
\label{eq:pulled}
\chi^2_{\rm pull}(\varepsilon)
= 
\frac{f^2}{\delta R_{\rm sys}^2}
+
\sum_\theta
\frac{
\left[
R^{\theta,\rm SM}
-
(1+f)R^{\theta,\rm NSI}(\varepsilon)
\right]^2
}{
\left(\delta R_{\rm stat}^{\theta} R^{\theta,\rm SM}\right)^2
} ,
\end{equation}
where we minimize over $f$ that represents a position-independent fractional normalization
shift in the predicted ratio.

Eq.~\eqref{eq:pulled} provides an idealized benchmark intended
to isolate the information supplied by detector motion. We do not
separately model uncertainties that can vary among positions 
including those associated with the flux spectral shape, interaction
and nuclear modeling, detector response, event selection  and
beam related backgrounds. The projected sensitivities we obtain therefore
illustrate the potential reach of the approach. We leave detailed modeling of these effects for dedicated experimental 
analyses.

\begin{figure*}[t]
    \centering
    \begin{minipage}{0.48\textwidth}
        \centering
        \includegraphics[width=\linewidth]{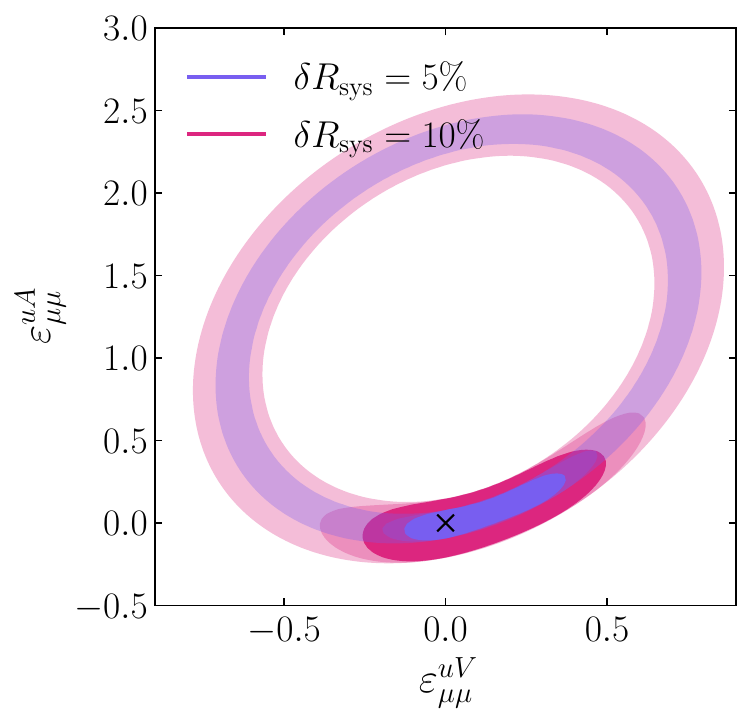}
        \includegraphics[width=\linewidth]{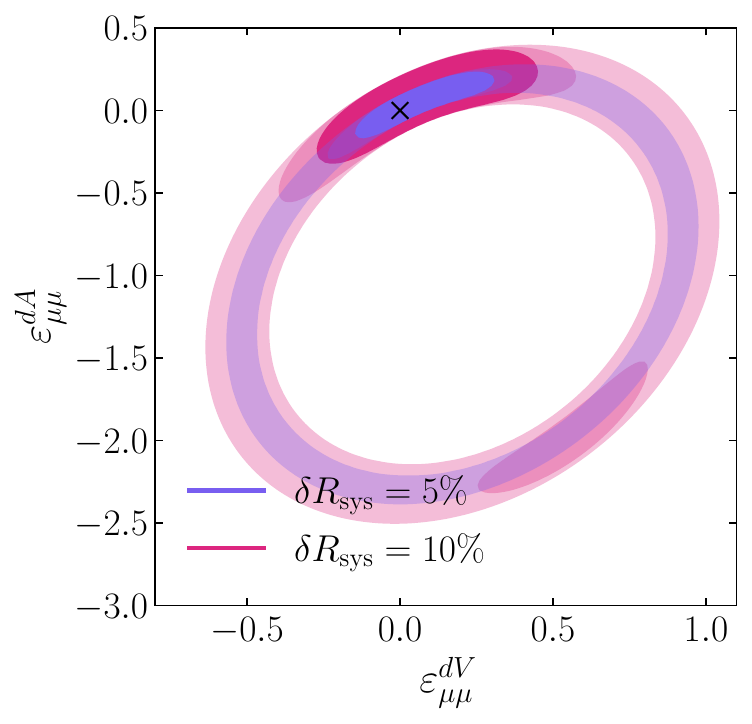}
    \end{minipage}
    \hfill
    \begin{minipage}{0.48\textwidth}
        \centering
        \includegraphics[width=\linewidth]{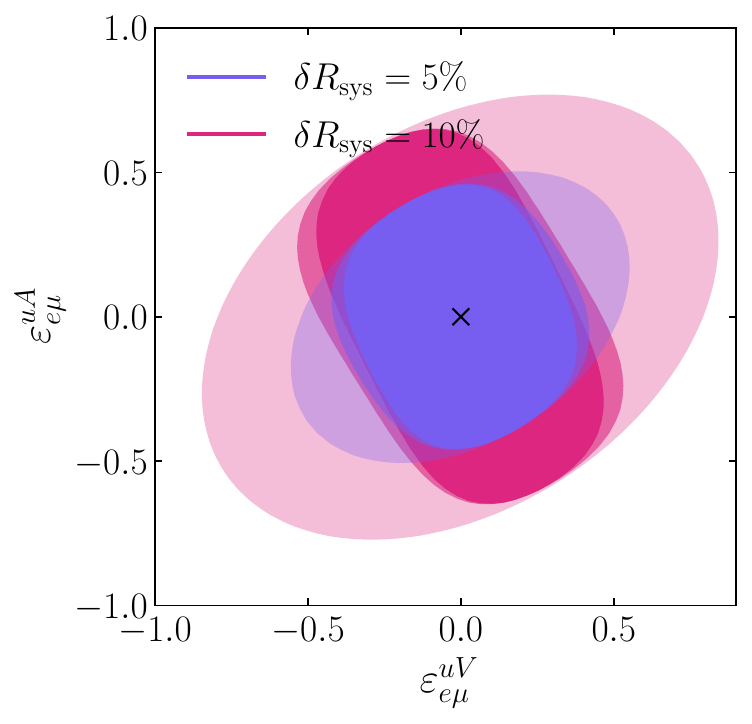}
        \includegraphics[width=\linewidth]{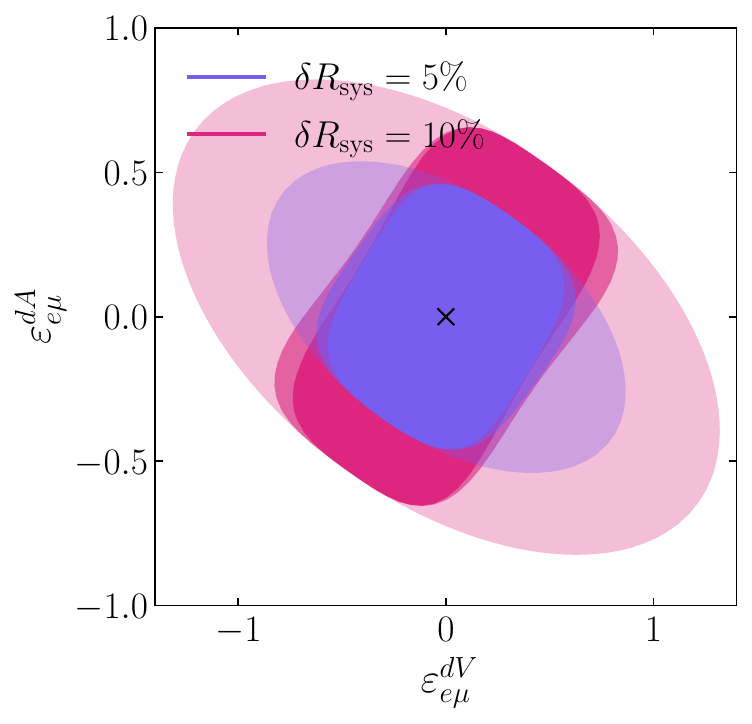}
    \end{minipage}
    \caption{ 
Sensitivity projection regions in the vector-axial NC NSI
parameter space obtained with $\chi_{\rm pull}^2$, considering $95\%$ confidence level. The upper and
lower panel rows show couplings to $u$ and $d$ quarks, respectively.
Flavor-diagonal coefficients are shown in the left panel column and
off-diagonal coefficients producing a $\nu_e$ final state are shown
in the right panel column. The two colors correspond to
$\delta R_{\rm sys}=5\%$ (blue) and $10\%$ (magenta). The black cross marks the SM regime. For each systematic uncertainty benchmark 
the regions correspond to one, two  and three off-axis angle detector position measurements,
considering the same total exposure divided equally among the included
positions. Within each color, the lightest, intermediate and darkest regions correspond to one, two  and three detector positions, respectively. A single integrated ratio leaves extended vector-axial
degeneracies, whereas multiple incident spectra constrain their
energy dependence and reduce the allowed regions.    
    }
    \label{fig:2d_chi2}
\end{figure*}

In Fig.~\ref{fig:chi2} we compare the two $\chi^2$ treatments. Considering  a single detector position 
NSI effects in the event rate ratio can be degenerate with
an overall normalization shift. However, combining measurements at multiple
off-axis positions adds information from the relative variation of
$R^{\theta}$ across the different incident spectra.

The improvement in sensitivity due to measurements at different positions is particularly significant for NC NSI. Vector and
axial contributions modify the energy and momentum transfer
dependence of the NCQE cross-section and therefore produce different
variations across detector positions. In contrast, the left-handed CC
coefficient $\varepsilon_{\mu\mu}^{udL}$ uniformly rescales the SM
CCQE cross-section and thus benefits significantly less from the
addition of multiple off-axis measurements.

For flavor-diagonal NC NSI  interference with the SM amplitude can
also generate secondary solutions for which the predicted ratio at a
single detector position is close to the SM value. Measurements using several
incident spectra at different detector locations can restrict and remove these solutions since the cancellation
does not generally persist at every off-axis position. On the other hand, off-diagonal
NC coefficients instead enter quadratically and therefore do not
provide sensitivity to the sign of a single real coefficient.
Multiple spectra can restrict or eliminate secondary solutions, although disconnected solutions with larger couplings can remain for particular  parameters.

For the one-parameter scans shown in Fig.~\ref{fig:chi2}, the projected
$95\%$ confidence sensitivity is defined by $ 
\Delta\chi^2=3.84 $ 
corresponding to one parameter  of interest. 
For larger systematic uncertainties, the correlated normalization pulled $\chi_{\rm pull}^2$
analysis benefits more significantly from combining multiple off-axis
angle detector positions, because the relative variation of $R^\theta$ across angles
remains sensitive to NSI even when the overall normalization is
allowed to vary. This additional spectral information also enables removing
degenerate solutions that can reproduce the SM rate at a single
position. 
For a single position the two treatments $\chi_{\rm base}^2$ and $\chi_{\rm pull}^2$ give numerically similar results in the vicinity of the SM prediction, where $R^{\rm NSI}  \simeq R^{\rm SM}$. To test robustness, we have repeated the $\varepsilon_{\mu\mu}^{uA}$ analysis using $M_A = 1.03$~GeV and $Q_{PB}^2 = 0.03$~GeV$^2$ and found that the sensitivity results were not significantly affected.

Off-diagonal vector NSI have also been considered in interpretations
of long baseline oscillation data, including comparisons of the T2K
and NOvA results
\cite{Chatterjee:2020kkm,Denton:2020uda,Chatterjee:2024kbn}.
In ordinary unpolarized matter, only vector NSI contribute to the
coherent matter potential whereas axial NSI do not.
With the conventions adopted here, the incident $\nu_\mu$ beam
at IWCD probes $\varepsilon_{e\mu}^{qV}$ and
$\varepsilon_{\tau\mu}^{qV}$. For Hermitian NC interactions, these
correspond, up to complex conjugation, to the same off-diagonal matrix
elements that enter neutrino oscillation propagation analyses. We take all coefficients to
be real in the present study.
The comparison between neutrino NSI scattering and oscillation sensitivities is
not direct. Oscillation experiments typically constrain  
combinations of the quark-level vector couplings weighted by the electron, up-quark  and down-quark densities in matter
and their allowed regions can depend on several flavor coefficients,
 phases  and correlations with the standard oscillation
parameters. Instead, IWCD enables probing the underlying NSI up-quark and down-quark
couplings directly through scattering. Its sensitivity
is therefore complementary to long baseline neutrino oscillation fits.

Next, we allow vector and axial coefficients to vary simultaneously.
For each quark flavor $q=\{u,d\}$, we consider the flavor-diagonal plane
$(\varepsilon_{\mu\mu}^{qV},\varepsilon_{\mu\mu}^{qA})$ and the
off-diagonal plane
$(\varepsilon_{e\mu}^{qV},\varepsilon_{e\mu}^{qA})$.
The projected $95\%$ confidence regions are defined by $ 
\Delta\chi^2=5.99$  
corresponding to two parameters of interest, with the common
normalization nuisance parameter $f$ profiled at each point.

In Fig.~\ref{fig:2d_chi2} we display the resulting fit contour regions. Noticeably, a single detector position
integrated event rate ratio leaves an extended degeneracy between
vector and axial contributions. 
Through the convolution with varying fluxes at different beam angles, the movable detector is able to restrict 
the allowed parameters to combinations that reproduce both measured
ratios, while a third position measured spectrum provides an additional independent
energy weighting and further reduces the allowed parameter space region.

With two detector positions considered in analysis, the branch of the flavor-diagonal NSI degeneracy opposite the SM point is found to be eliminated for up-quark couplings. For down-quark couplings, it is substantially reduced for 10\% systematic uncertainty and eliminated for 5\%. 
Inclusion of a third detector position  in analysis removes this branch in both cases and further contracts the region connected to the SM point. For off-diagonal NSI, additional spectra primarily rotate and shrink the allowed regions since only the vector and axial new physics amplitudes interfere with one another.

 \begin{figure*}[t]
    \centering
    \begin{minipage}{0.48\textwidth}
        \centering
        \includegraphics[width=1\linewidth]{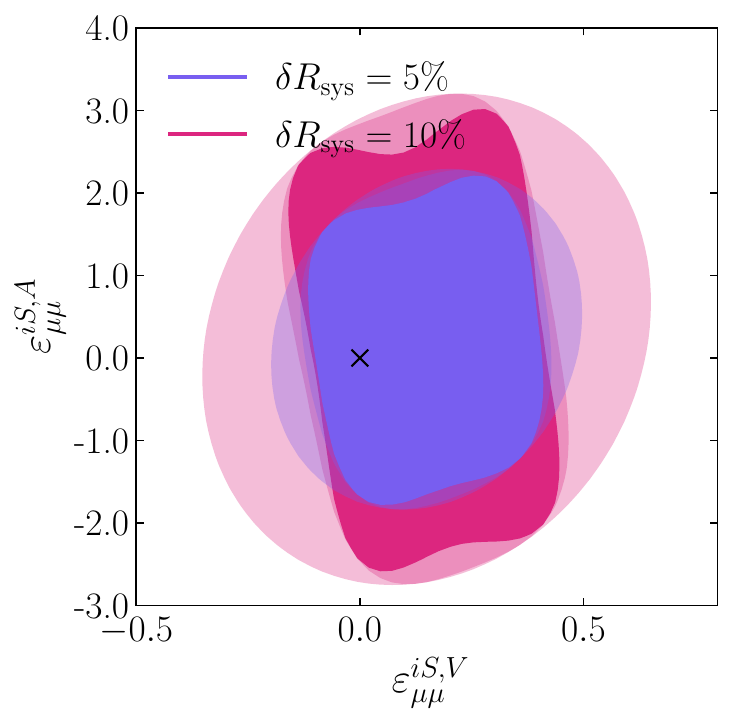}  
        \includegraphics[width=1\linewidth]{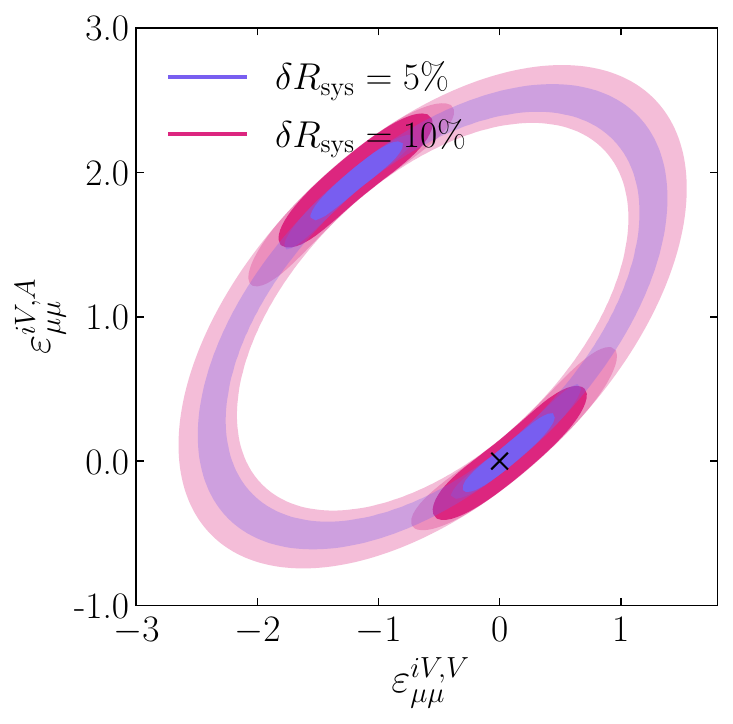} 
    \end{minipage}
    \hfill
    \begin{minipage}{0.48\textwidth}
        \centering
        \includegraphics[width=1\linewidth]{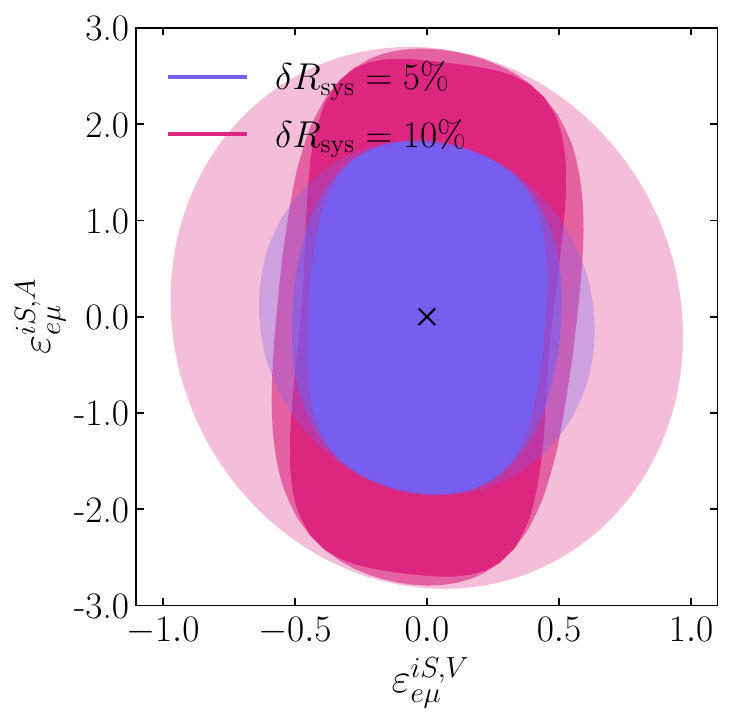}
        \includegraphics[width=1\linewidth]{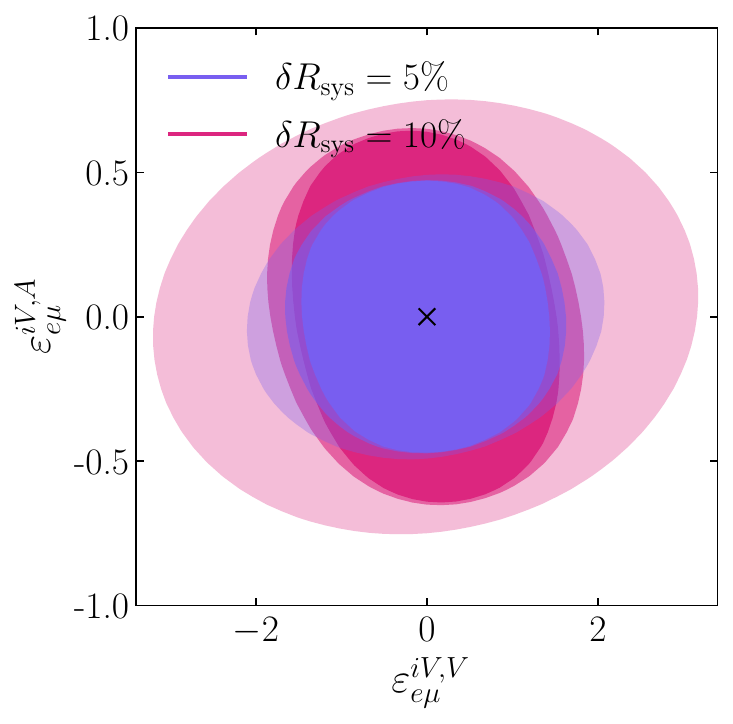}
    \end{minipage}
    \caption{Same as Fig.~\ref{fig:2d_chi2},  but in the isoscalar-isovector basis. The black cross marks the SM regime point. Within each color, the lightest, intermediate and darkest regions correspond to one, two  and three detector positions, respectively. }
    \label{fig:2d_chi2_isospin}
\end{figure*}

For flavor-diagonal NSI both SM-NSI contribution interference and interference
between vector and axial contributions determine the contour shapes.
For off-diagonal NSI  the vector and axial amplitudes do not interfere
with the SM since they produce a different final state neutrino
flavor. However, these contributions can interfere with one another when they
produce the same final state flavor, which allows cancellations within the
pure NSI contribution.

The IWCD sensitivity projections are complementary to constraints from
higher energy neutrino scattering experiments such as CHARM
\cite{CHARM:1987pwr}, NuTeV~\cite{NuTeV:2001whx}, and to projected measurements at FASER$\nu$
\cite{Ismail:2020yqc,Kling:2025lnt}. These experiments probe related
operators primarily through deep inelastic scattering at substantially
higher energies, whereas IWCD probes  interactions on an
oxygen target at sub-GeV to few-GeV energies. Direct comparison depends on
operator basis, flavor assumptions  and confidence prescriptions. We
therefore leave combined NSI fit analyses for future work.

 \begin{figure*}[t] 
    \centering
    \includegraphics[width=0.48\linewidth]{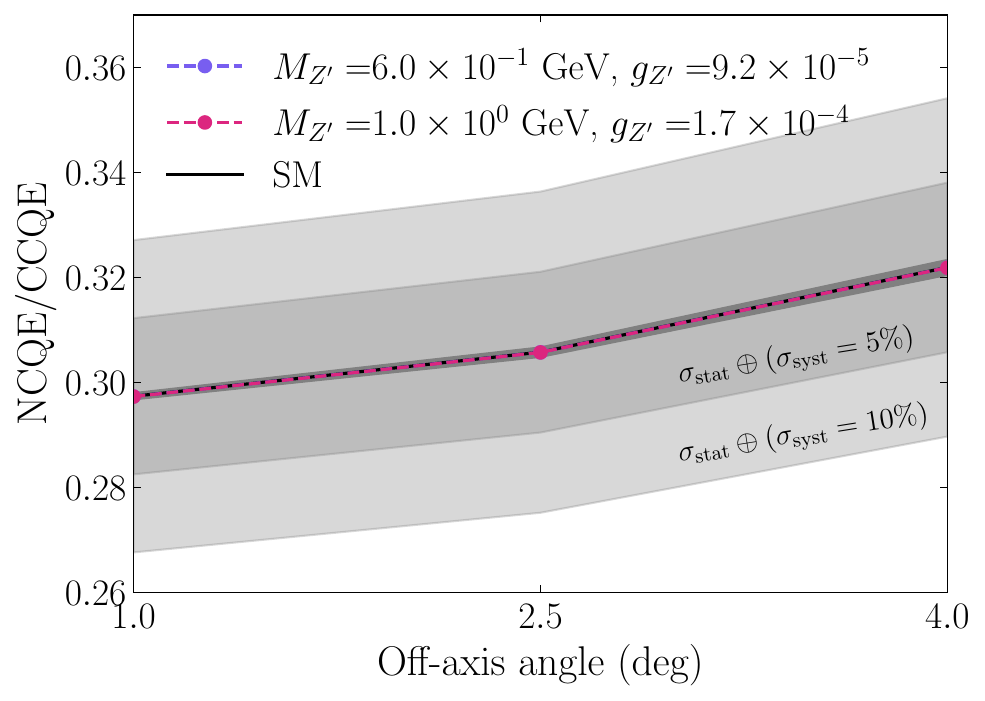}
    \includegraphics[width=0.48\linewidth]{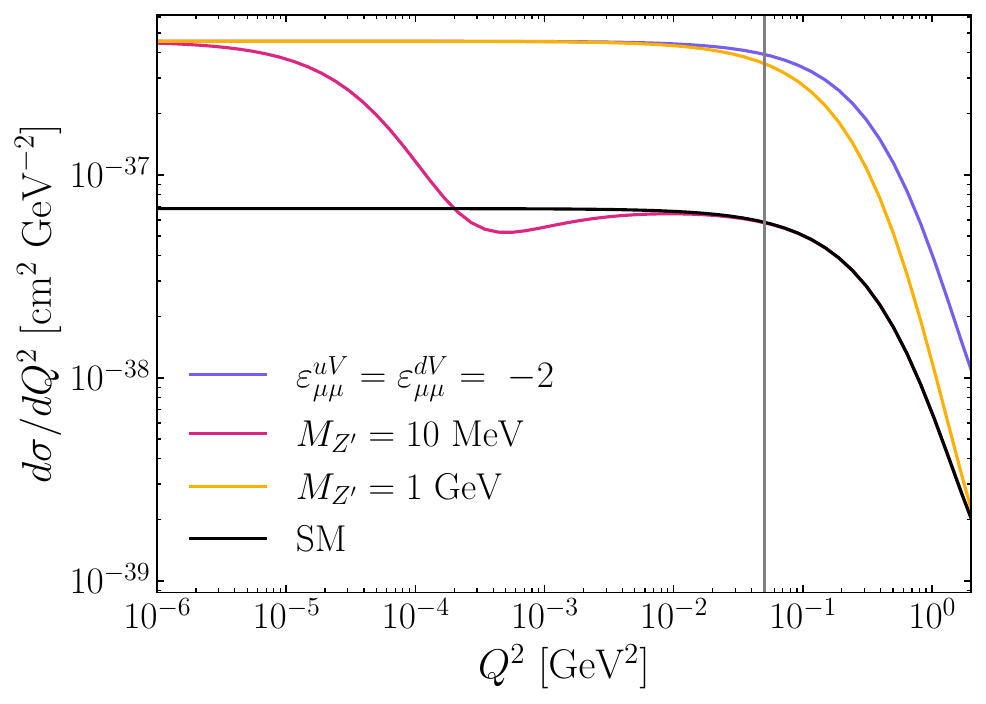}
    \caption{(Left) NCQE to CCQE event rate ratio for representative light vector 
mediator NSI benchmarks. The same  $Q_{\rm PB}^2$ cutoff as in the main analysis is applied.  The NSI couplings are
chosen as representative examples. (Right) 
Differential NCQE cross-section for $E_\nu=1~{\rm GeV}$ without low $Q_{\rm PB}^2$ cut,
including vector NSI mediators with masses $M_{Z'}=10~{\rm MeV}$
and $1~{\rm GeV}$. The couplings are chosen for illustration to reproduce the
 low energy contact coefficients
$\varepsilon_{\mu\mu}^{uV}=\varepsilon_{\mu\mu}^{dV}=-2$.
Such deliberately large values are used only to showcase the
mediator mass dependence and are not employed in the
sensitivity analyses. The vertical line denotes the
phenomenological benchmark cutoff $Q_{\rm PB}^2$ adopted in the main analysis, such that only the region $Q^2 \geq Q_{\rm PB}^2$ is retained for event rate computations.} 
    \label{fig:light_med}
\end{figure*}

Light NSI mediators are considered in
App.~\ref{app:lightmed}. For the benchmark low $Q^2$  
suppression adopted in our analysis, a significant portion of the light mediator enhancement regime lies
below the accessible momentum transfer region. In such case, detector motion and multiple measurements provide primarily overall
event rates.

\section{Conclusions}\label{sec:conclusion}

We have shown that a movable neutrino detector can utilize measurements at
multiple off-axis positions to probe the structure of fundamental
interactions. At the movable IWCD, the same target and detector are exposed to
distinct incident energy spectra at different detector positions while many experimental and
interaction model uncertainties remain correlated. The resulting
variation of the NCQE-CCQE event rate ratio across positions
provides information that is not available from a single integrated
measurement.

We demonstrated this strategy for probing neutrino NSI.
Combining three off-axis detector positions and adopting a 5\% correlated normalization uncertainty benchmark we found  95\% CL projected sensitivities in the heavy mediator scenario to axial neutral current NSI of $-0.07 \lesssim \varepsilon^{uA}_{\mu\mu} \lesssim 0.06$ and vector NSI of $-0.10 \lesssim \varepsilon^{uV}_{\mu\mu} \lesssim 0.12$, as well as for charged current NSI of $-0.05 \lesssim \varepsilon^{udL}_{\mu\mu} \lesssim 0.05$. Flavor-diagonal vector and axial NC interactions modify the energy and
momentum transfer dependence of the NCQE cross section differently, thus
combining several off-axis spectra can reduce normalization and
vector-axial degeneracies. In contrast, the left-handed CC
coefficient $\varepsilon_{\mu\mu}^{udL}$ predominantly rescales the
SM CCQE cross section and therefore gains less from multiple detector position measurements.
Off-diagonal NC interactions can also be probed directly, although
their leading contribution is quadratic since they do not interfere
with the SM amplitude.
 
The projected IWCD NSI sensitivities are complementary to both neutrino oscillation
and higher energy neutrino scattering measurements. Oscillation experiments
constrain combinations of vector NSI weighted by matter distribution, whereas IWCD directly
probes the underlying $u$-quark and $d$-quark interactions through
scattering. In particular, axial NSI do not contribute to
the coherent matter potential in ordinary matter as relevant for neutrino propagation effects. The approach we outline is complementary to measurements with other techniques that 
test different combinations of the same
underlying interactions. Our work further calls for dedicated and combined analyses. For the light mediator NSI benchmarks considered here, we identify
 improved access to the low momentum transfer parameter regions as a key lever
for enhancing the spectral sensitivity enabled by detector motion.

More broadly, we have demonstrated that measurements at multiple detector  positions can turn movable
detectors into sensitive spectroscopic probes of new physics. The strategy is
applicable whenever interaction
structures produce distinct spectral responses  and can be extended
to other processes, beam configurations  and movable or
multi-position detector concepts.

\section*{Acknowledgments}
We thank Megan Friend, Mark Hartz, Akira Konaka
 and Patrick De Perio for useful discussions.
T.S. and V.T. were supported by the World Premier International Research Center Initiative (WPI), MEXT, Japan. 

\appendix

\section{Isoscalar and Isovector Combinations}\label{app:isospin}

Isoscalar and isovector combinations enable convenient parameterization of NSI in terms of isospin, and are useful for organizing scattering on approximately isoscalar oxygen. Analyses focusing on neutrino oscillation matter effects instead constrain density weighted combinations of electron, up-quark and down-quark vector NSI. For approximately isoscalar terrestrial matter  the quark contribution is closely related to, but not identical with  the isoscalar combination defined below. 

In Fig.~\ref{fig:2d_chi2_isospin}   we display our results recast in this parameter space, considering
\begin{align} \label{eq:isonsi}
    \varepsilon_{\alpha\beta}^{iS,X} &= \varepsilon_{\alpha\beta}^{uX} + \varepsilon_{\alpha\beta}^{dX} \nonumber \\
    \varepsilon_{\alpha\beta}^{iV,X} &= \varepsilon_{\alpha\beta}^{uX} - \varepsilon_{\alpha\beta}^{dX}~,
\end{align}
where $X = \{A, V \}$. We do not include additional 1/2 factor in these definitions.

\section{Light Mediator Non-Standard Interactions}
\label{app:lightmed}

The contact interaction NSI framework considered in the main
analysis applies in the heavy mediator limit where
$m_{\rm med}^2 \gg Q^2$, where the mediator propagator is
approximately momentum independent. If the mediator mass is
comparable to or smaller than the characteristic momentum
transfer its propagator must instead be retained and produces an
additional $Q^2$ dependence in the NC amplitude.

We consider a vector mediator $Z'$ with interactions
\begin{align}
    \mathcal{L}
    \supset&
    \ g_{\nu} 
    \overline{\nu}_{\mu}\gamma^\rho P_L\nu_{\mu} Z'_\rho
    + \sum_{q=u,d}
    g_q \overline{q}\gamma^\rho q Z'_\rho ,
    \label{eq:lightlag}
\end{align}
where $g_{\nu}$ and $g_q$ are couplings. These interactions result in vector NSI.
Matching this interaction onto the vector NSI convention of
Eq.~\eqref{eq:VA_nsi} gives the
momentum dependent coefficient~\cite{AtzoriCorona:2022moj}
\begin{align}
    \varepsilon^{qV}_{\mu\mu}(Q^2)
    =
    \frac{g_{\nu}g_q}
    {\sqrt{2}G_F\left(Q^2+M_{Z'}^2\right)} ,
    \label{eq:lightmednsi}
\end{align}
where $Q^2 >0$. The sign of the product
$g_{\nu}g_q$ determines whether the new amplitude interferes
constructively or destructively with the SM contribution. 

As a concrete model realization we consider $U(1)_{B - 3 L_{\mu}}$~\cite{Heeck:2018nzc,AtzoriCorona:2022moj}, which is anomaly free once the SM is extended by right handed neutrinos. The relevant couplings are
\begin{equation}
    g_{\nu_{\mu}} = -3 g_{Z'}~~~,~~~g_u = g_d = \dfrac{g_{Z'}}{3}~.
\end{equation}  
Then, substituting into Eq.~\eqref{eq:lightmednsi}
\begin{align}
    \varepsilon^{uV}_{\mu\mu}(Q^2) = \varepsilon^{dV}_{\mu\mu}(Q^2)
    =
    - \frac{g_{Z'}^2}
{\sqrt{2}G_F\left(Q^2+M_{Z'}^2\right)} .
    \label{eq:lightmednsimod}
\end{align}
The sign affects the interference with SM contributions. In other model implementations, however, opposite sign and different interference with SM contributions are possible and can modify sensitivity projections as well as constraints. 

We implement Eq.~\eqref{eq:lightmednsimod} through the same
modifications of the nucleon form factors used for the
heavy mediator NSI in Eq.~\eqref{eq:NSI_ff}. The full
cross-section  includes both the SM-$Z'$
interference contribution  proportional to
$(Q^2+M_{Z'}^2)^{-1}$  and the purely mediator contribution,
proportional to $ (Q^2+M_{Z'}^2)^{-2}$.

Fig.~\ref{fig:light_med} shows the resulting
NCQE to CCQE event rate ratio at the three benchmark IWCD
positions. Although the propagator introduces additional
momentum dependence  for the benchmark couplings and the
low $Q^2$ treatment adopted here its effect is largely similar
across the three incident spectra. The resulting ratios
therefore exhibit only a weak additional dependence on the
off-axis detector position. Thus, for the benchmark positions considered the light
mediator predominantly modifies the overall NCQE rate rather
than producing distinct position dependent features.

Due to momentum transfer dependence the sensitivity to light mediator NSI effects is governed by the
lowest accessible momentum transfers. In our benchmark
treatment we impose
$Q^2 \geq Q_{\rm PB}^2 \simeq 0.05~{\rm GeV}^2$.
Consequently, for
$M_{Z'}^2 \ll Q_{\rm PB}^2$  the accepted contributing events lie primarily
in the regime $Q^2\gg M_{Z'}^2$  where the propagator approaches
$1/Q^2$ behavior.  The low mass sensitivity then becomes approximately
independent of $M_{Z'}$, while the strongest low $Q^2$
enhancement is removed by the imposed cutoff. This behavior is shown in Fig.~\ref{fig:light_med} for example couplings that are deliberately chosen to be enhanced for illustration, and we do not employ them in our sensitivity analyses.  We emphasize that
$Q_{\rm PB}^2$ is a phenomenological representation of nuclear
suppression rather than an exact kinematic threshold, a
realistic nuclear treatment can allow contributions below this
nominal value. 

\begin{figure}[t] 
    \centering
    \includegraphics[width=\linewidth]{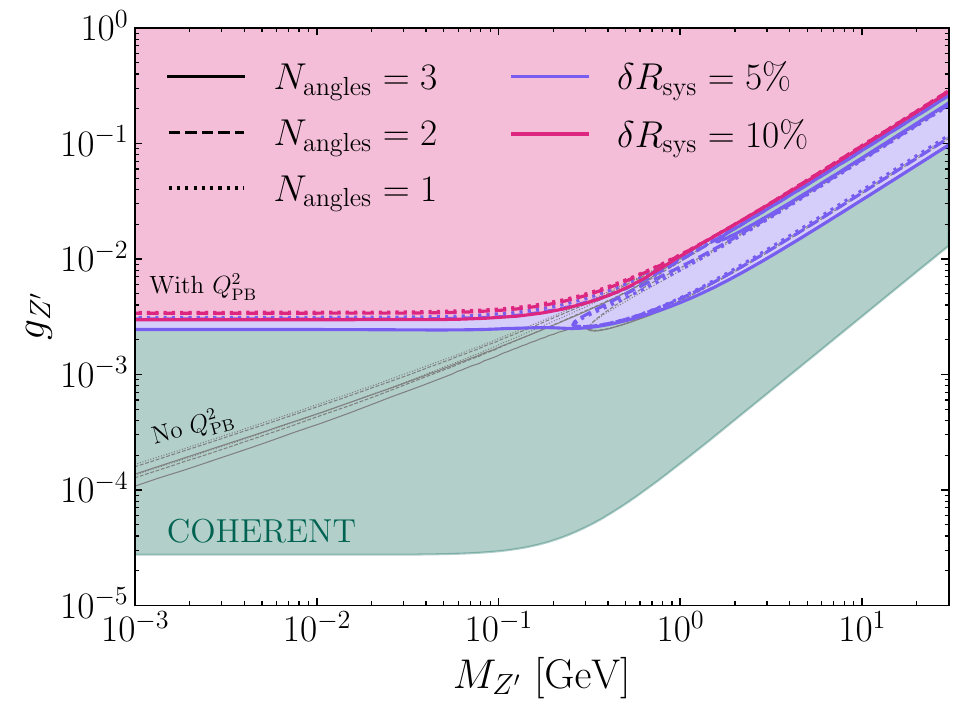}
    \caption{Sensitivity projections at 95\% confidence level on the coupling of a light
vector mediator NSI as a function of mediator mass. For the couplings, we assume $(B - 3 L_{\mu})$ model.
The red (blue) shaded regions show the sensitivity with measurements at three off-axis detector positions while the dashed and dotted lines show two and one position respectively.  
The shaded regions include the phenomenological  $Q_{\rm PB}^2$ cutoff adopted in the main analysis.
Thin curves in this figure denote results obtained without the phenomenological $Q_{\rm PB}^2$ cutoff. The green shaded region is excluded at $2\sigma$-level by the combined COHERENT CsI+Ar analysis for the $(B - 3 L_{\mu})$ model~\cite{DeRomeri:2022twg,AtzoriCorona:2022moj}.}
    \label{fig:light_med_chi2}
\end{figure}

At each fixed mediator mass we apply the pulled
$\chi^2$ analysis of Eq.~\eqref{eq:pulled} to one, two and
three off-axis positions and derive nominal 95\% confidence level sensitivity
projections on the effective coupling $g_{Z'}$. The results are displayed
in Fig.~\ref{fig:light_med_chi2}. With the benchmark
low $Q^2$ suppression included the sensitivities from the
different position configurations are found to be similar, consistent with
the weak position dependence in Fig.~\ref{fig:light_med}.
For comparison, the thin curves show the result obtained
without the phenomenological $Q_{\rm PB}^2$ cutoff and
illustrate the significance of the low $Q^2$ response for light
mediator NSI searches.

In Fig.~\ref{fig:light_med_chi2} we further overlay representative light mediator NSI constraints
from COHERENT experiment, assuming  the $(B - 3 L_{\mu})$ coupling
benchmark of Refs.~\cite{DeRomeri:2022twg,AtzoriCorona:2022moj}.
The IWCD projections and the COHERENT constraints use the same coupling assignment. The overlay is representative rather than a complete  landscape of constraints and searches. The projections shown
here are phenomenological benchmarks demonstrating that improved control and acceptance of the
low momentum transfer region play an important role for exploiting
detector motion in light mediator NSI searches.
 
\bibliography{references}

\end{document}